\documentclass[a4paper,11pt]{article}%
\usepackage{jheppub}
\usepackage{graphicx}
\usepackage{amsmath}
\usepackage{amsfonts}
\usepackage{amssymb}%
\providecommand{\U}[1]{\protect\rule{.1in}{.1in}}
\affiliation{Budker Institute of Nuclear Physics and Novosibirsk State University,\\630090 Novosibirsk, Russia}
\emailAdd{A.V.Grabovsky@inp.nsk.su}
\abstract{The connected contribution to the kernel of the evolution equation for the 3-quark Wilson loop operator was derived within Balitsky high energy operator expansion. Its C-odd part was linearized and transferred to the momentum space.}
\arxivnumber{}
\begin{document}

\title{\boldmath Connected contribution to the kernel of the evolution equation for
3-quark Wilson loop operator}
\author{A. V. Grabovsky}
\maketitle

\flushbottom

\section{Introduction}

The Balitsky-Fadin-Kuraev-Lipatov (BFKL) equation \cite{Fadin:1975cb}%
--\cite{Balitsky:1978ic} governs the energy evolution of the pomeron Green
function. Pomeron is the C-even bound state of two reggeized gluons whereas
its C-odd counterpart consisting of three reggeized gluons is known as
odderon. The evolution equation for odderon Green function is the
Bartels-Kwiecinski-Praszalowicz (BKP) equation \cite{Bartels:1980pe}%
-\cite{Kwiecinski:1980wb}. While the next to leading order (NLO) corrections
to the kernel of the BFKL equation have been known for some time
\cite{Fadin:1998py}-\cite{Fadin:2005zj}, the nonforward NLO BKP\ kernel for
odderon exchange has been calculated only quite recently \cite{Bartels:2012sw}%
. It consists of 3 pairwise octet kernels and a connected $3\rightarrow3$ contribution.

An alternative approach to the Regge limit of high energy QCD is based on the
Balitsky-Kovchegov (BK) equation \cite{Balitsky}--\cite{Kovchegov}. Its NLO
form was found in \cite{Balitsky:2006wa}-\cite{Balitsky:2008zza}. The
derivation of the BK equation given by I. Balitsky in \cite{Balitsky} through
Wilson line technique does not assume reggeization. However, the linear part
of the BK equation coincides with the so called Moebius form of the BFKL
equation valid for scattering of colorless particles in the linear regime
\cite{Fadin:2006ha}. In NLO these kernels coincide after an equivalence
transformation \cite{Fadin:2009gh}, which changes the kernel without changing
the observables.

The BK Green function is a color dipole. However, in the C-odd case it is not
the most general Green function since it depends only on 2 coordinates, while
odderon consists of 3 reggeized gluons. The 3-quark Wilson loop (3QWL) is
another colorless operator which has a baryon structure $\varepsilon
^{i^{\prime}j^{\prime}h^{\prime}}\varepsilon_{ijh}U_{1i^{\prime}}%
^{i}U_{2j^{\prime}}^{j}U_{3h^{\prime}}^{h}$. Its linear evolution equation was
proved equivalent to the C-odd BKP one in \cite{Hatta:2005as} and its
nonlinear evolution equation was derived in \cite{Gerasimov:2012bj}. In the
momentum representation the evolution of this operator was studied in
\cite{Praszalowicz:1997nf}, \cite{Braunewell:2005ct}\ and the nonlinear
equation was worked out in \cite{Bartels:2007aa}.

There is a prompt question of the NLO\ kernel for the 3QWL operator. In this
paper the connected contribution to such a kernel has been calculated within
Balitsky high energy operator expansion \cite{Balitsky:2008zza}. The linear
part of this contribution for the C-odd case was transferred to the momentum
representation and found to be different from the connected $3\rightarrow3$
kernel of \cite{Bartels:2012sw}. It indicates that there should be an
equivalence transformation connecting the kernels.

In this paper the dimension of the space-time is kept equal to 4 since the
connected part of the NLO kernel does not contain the UV divergencies and the
sum of the diagrams is IR stable because the 3QWL is a gauge invariant
colorless operator.

The paper is organized as follows. The next section contains the definitions
and the derivation of the leading order (LO) evolution equation for the 3QWL
operator. Section 3 deals with the connected contribution proportional to the
second iteration of the LO kernel. Section 4 presents the calculation of the
connected contribution with 2 gluon intersections of the shockwave. Section 5
comprises the calculation of the diagrams with 1 gluon intersection of the
shockwave. Section 6 gives the Furier transform of the linearized result for
the C-odd case. The last section concludes the paper.

\section{Definitions and necessary results}

We introduce the light cone vectors $n_{1}$ and $n_{2}$%
\begin{equation}
n_{1}=\left(  1,0,0,1\right)  ,\quad n_{2}=\frac{1}{2}\left(  1,0,0,-1\right)
,\quad n_{1}^{+}=n_{2}^{-}=n_{1}n_{2}=1,
\end{equation}
and for any vector $p$ we have%
\begin{equation}
p^{+}=p_{-}=pn_{2}=\frac{1}{2}\left(  p^{0}+p^{3}\right)  ,\qquad p_{+}%
=p^{-}=pn_{1}=p^{0}-p^{3},
\end{equation}%
\begin{equation}
p=p^{+}n_{1}+p^{-}n_{2}+p_{\bot},\qquad p^{2}=2p^{+}p^{-}-\vec{p}^{\,2},
\end{equation}%
\begin{equation}
\quad p\,k=p^{\mu}k_{\mu}=p^{+}k^{-}+p^{-}k^{+}-\vec{p}\vec{k}=p_{+}%
k_{-}+p_{-}k_{+}-\vec{p}\vec{k}.
\end{equation}
We work in the light-cone gauge $\mathcal{A}n_{2}=0$ and in our convention the
3-gluon interaction Lagrangian has the form%
\begin{equation}
\mathcal{L}_{i}\mathcal{=}-gf^{abc}\left(  \partial_{\mu}\mathcal{A}_{\nu}%
^{a}\right)  \mathcal{A}^{b\mu}\mathcal{A}^{c\nu}.
\end{equation}
We would like to calculate the connected part of the kernel for the evolution
equation for the 3-quark Wilson loop operator%
\begin{equation}
B_{123}^{\eta}=\varepsilon^{i^{\prime}j^{\prime}h^{\prime}}\varepsilon
_{ijh}U\left(  \vec{z}_{1},\eta\right)  _{i^{\prime}}^{i}U\left(  \vec{z}%
_{2},\eta\right)  _{j^{\prime}}^{j}U\left(  \vec{z}_{3},\eta\right)
_{h^{\prime}}^{h} \label{B}%
\end{equation}
contributing to the evolution of a baryon Green function. Hereafter we will
use the following shorthand notation for such convolutions%
\begin{equation}
\varepsilon^{i^{\prime}j^{\prime}h^{\prime}}\varepsilon_{ijh}U\left(  \vec
{z}_{1},\eta\right)  _{i^{\prime}}^{i}U\left(  \vec{z}_{2},\eta\right)
_{j^{\prime}}^{j}U\left(  \vec{z}_{3},\eta\right)  _{h^{\prime}}^{h}%
=U_{1}\cdot U_{2}\cdot U_{3},
\end{equation}
where%
\begin{equation}
U\left(  \vec{z},\eta\right)  =Pe^{ig\int_{-\infty}^{+\infty}b_{\eta}%
^{-}(z^{+},\vec{z})dz^{+}} \label{WL}%
\end{equation}
is the Wilson line with the path along the $z^{-}=0$ line and $b_{\eta}^{-}$
is the external shock-wave field built from only slow gluons
\begin{equation}
b_{\eta}^{-}=\int\frac{d^{4}p}{\left(  2\pi\right)  ^{4}}e^{-ipz}b^{-}\left(
p\right)  \theta(e^{\eta}-|p^{+}|).
\end{equation}
The parameter $\eta$ separates the slow gluons entering the Wilson lines from
the fast ones in the impact factors. The shape of the path at $z^{+}=\pm
\infty$ in (\ref{WL}) is not important because the field is concentrated at
$z^{+}=0.$ The gluon field consists of the fast component $A$ with the
rapidities greater than $\eta$ and the slow one $b_{\eta}^{-}$
\begin{equation}
\mathcal{A}=A+b,\qquad b^{\mu}\left(  z\right)  =b^{-}(z^{+},\vec{z}%
)n_{2}^{\mu}=\delta(z^{+})b\left(  \vec{z}\right)  n_{2}^{\mu}.
\end{equation}
To derive the evolution equation one has to calculate the operator
$B_{123}^{\eta}$\ in the shockwave background $\langle\rangle$, i.e. integrate
it over the gluons with $\sigma=e^{\eta}>p^{+}>\sigma_{1},$ where $\sigma
_{1}\ll\sigma$ is the lower cutoff set by the target
\begin{equation}
\langle B_{123}^{\eta}\rangle=\frac{\langle0|T(B_{123}^{\eta}e^{i\int
\mathcal{L}\left(  z\right)  dz})|0\rangle}{\langle0|T(e^{i\int\mathcal{L}%
\left(  z\right)  dz})|0\rangle}. \label{<B>}%
\end{equation}
To this end we need the gluon propagator in the light cone gauge. The free
gluon propagator reads%
\begin{equation}
G_{0}^{\mu\nu}\left(  p\right)  =\frac{-id^{\mu\nu}\left(  p\right)  }%
{p^{2}+i0},
\end{equation}
with
\begin{equation}
d^{\mu\nu}\left(  p\right)  =g^{\mu\nu}-\frac{p^{\mu}n_{2}^{\nu}+p^{\nu}%
n_{2}^{\mu}}{pn_{2}}=g_{\bot}^{\mu\nu}-\frac{p_{\bot}^{\mu}n_{2}^{\nu}%
+p_{\bot}^{\nu}n_{2}^{\mu}}{p^{+}}-2\frac{n_{2}^{\mu}n_{2}^{\nu}p^{-}}{p^{+}},
\end{equation}
Then%
\begin{equation}
G_{0}^{-j}(p^{+},x^{+},\vec{x})=-\frac{ix_{\bot}^{j}e^{i\frac{(\vec{x}%
^{\,\,2}+i0)p^{+}}{2x^{+}}}}{4\pi(x^{+})^{2}}\left(  \theta(x^{+})\theta
(p^{+})-\theta(-x^{+})\theta(-p^{+})\right)  ,
\end{equation}%
\begin{equation}
G_{0}^{--}(p^{+},x^{+},\vec{x})=\int\frac{d\vec{p}dp^{-}}{(2\pi)^{3}%
}e^{-ip^{-}x^{+}+i\,\vec{p}\,\vec{x}}\frac{2ip^{-}}{p^{+}(p^{2}+i0)}.
\label{G--}%
\end{equation}
One can take this integral explicitly. However, it is not convenient for us
since it introduces $\frac{1}{(p^{+})^{2}}$ singularity. Therefore we use
(\ref{G--}) for $G_{0}^{--}$ and integrate with respect to $x^{+}$ first. For
the calculation we need the following integral with $G_{0}^{--}$
\[
\int_{0}^{\infty}dz^{+}\int\frac{d\vec{p}dp^{-}}{(2\pi)^{3}}e^{-i(p^{-}%
-i\varepsilon)z^{+}+ip^{-}x^{+}+i\,\vec{p}(\vec{z}-\,\vec{x})}\frac
{2ip^{-}\theta(x^{+})\theta(p^{+})}{p^{+}(p^{2}+i0)}%
\]%
\begin{equation}
=\int\frac{d\vec{p}dp^{-}}{(2\pi)^{3}}\frac{e^{ip^{-}x^{+}+i\,\vec{p}(\vec
{z}-\,\vec{x})}2p^{-}\theta(x^{+})\theta(p^{+})}{p^{+}(p^{-}-i\varepsilon
)(2p^{+}p^{-}-\vec{p}^{\,\,2}+i0)}=0. \label{G--integral1=0}%
\end{equation}
The propagator in the shock-wave background field has the following two
convenient representations which we use in this paper
\[
G_{\mu\nu}^{ab}(x,y)|_{x^{+}>0>y^{+}}=-\int\frac{\theta\left(  p^{+}\right)
dp^{+}}{\left(  2\pi\right)  ^{3}}\frac{p^{+}}{2x^{+}y^{+}}\int d\vec
{z}e^{-ip^{+}\left\{  x^{-}-y^{-}+\frac{(\vec{z}-\vec{y})^{2}+i0}{2y^{+}%
}-\frac{(\vec{x}-\vec{z})^{2}+i0}{2x^{+}}\right\}  }%
\]%
\begin{equation}
\times\frac{g_{\bot\mu}^{\,\,\,\,\,\,\alpha}x^{+}-(x-z)_{\bot}^{\alpha}%
n_{2\mu}}{x^{+}}U_{\vec{z}}^{ab}\frac{g_{\bot\alpha\nu}(-y^{+})-(z-y)_{\bot
\alpha}n_{2\nu}}{-y^{+}}, \label{gluon_prop_through_exponontial}%
\end{equation}%
\[
G_{\mu\nu}^{ab}(x,y)|_{x^{+}>0>y^{+}}=\int\frac{\theta\left(  p^{+}\right)
dp^{+}}{\left(  2\pi\right)  ^{2}}\frac{i}{2x^{+}}\int d\vec{z}e^{-ip^{+}%
\left\{  x^{-}-y^{-}-\frac{(\vec{x}-\vec{z})^{2}+i0}{2x^{+}}\right\}  }%
\]%
\begin{equation}
\times\frac{g_{\bot\mu}^{\,\,\,\,\,\,\alpha}x^{+}-(x-z)_{\bot}^{\alpha}%
n_{2\mu}}{x^{+}}U_{\vec{z}}^{ab}\int\frac{d\vec{k}}{\left(  2\pi\right)  ^{2}%
}e^{i\vec{k}(\vec{z}-\vec{y})}e^{i\frac{y^{+}}{2p^{+}}\vec{k}^{2}}%
\frac{g_{\bot\alpha\nu}p^{+}-n_{2\nu}k_{\bot\alpha}}{p^{+}}. \label{GinSWmom}%
\end{equation}
The operator $\langle B_{123}^{\eta}\rangle$ in the shockwave background has
virtual the $B_{v}$ and the real $B_{r}$ contributions
\begin{equation}
\langle B_{123}^{\eta}\rangle=B_{v}+B_{r}. \label{Bv+Br}%
\end{equation}
One can find the real contribution using (\ref{gluon_prop_through_exponontial}%
) and integrating with respect to $z_{1}^{+}$ and $z_{2}^{+}$. The real
contribution from the interaction of $U_{1}$ and $U_{2}$ with the shockwave
reads
\[
B_{r12}=-g^{2}(U_{1}t^{a})\cdot(t^{b}U_{2})\cdot U_{3}\int_{-\infty}^{0}%
dz_{1}^{+}\int_{0}^{\infty}dz_{2}^{+}G^{--}(z_{2},z_{1})^{ba}%
\]%
\[
-g^{2}(t^{b}U_{1})\cdot\left(  U_{2}t^{a}\right)  \cdot U_{3}\int_{-\infty
}^{0}dz_{2}^{+}\int_{0}^{\infty}dz_{1}^{+}G^{--}(z_{1},z_{2})^{ba}%
\]%
\begin{equation}
=\frac{\alpha_{s}}{\pi^{2}}\int_{\sigma_{1}}^{\sigma}\frac{dp^{+}}{p^{+}}\int
d\vec{z}_{4}\frac{\left(  \vec{z}_{41}\vec{z}_{42}\right)  }{\vec{z}%
_{41}^{\,\,2}\vec{z}_{42}^{\,\,2}}U_{4}^{ba}\left(  (U_{1}t^{a})\cdot
(t^{b}U_{2})+(t^{b}U_{1})\cdot\left(  U_{2}t^{a}\right)  \right)  \cdot U_{3}.
\label{Br}%
\end{equation}
Similarly, the real contribution from the interaction of $U_{1}$ and the
shockwave reads
\[
B_{r1}=-g^{2}(t^{b}U_{1}t^{a})\cdot U_{2}\cdot U_{3}\int_{-\infty}^{0}%
dz_{1}^{\prime+}\int_{0}^{\infty}dz_{1}^{+}G^{--}(z_{1},z_{1}^{\prime})^{ba}%
\]%
\begin{equation}
=\frac{\alpha_{s}}{\pi^{2}}\int_{\sigma_{1}}^{\sigma}\frac{dp^{+}}{p^{+}}%
\int\frac{d\vec{z}_{4}}{\vec{z}_{41}^{\,\,2}}U_{4}^{ba}(t^{b}U_{1}t^{a})\cdot
U_{2}\cdot U_{3}. \label{Br1}%
\end{equation}
The virtual contribution from the $U_{1}$ and $U_{2}$ interaction reads%
\[
B_{v12}=-g^{2}(U_{1}t^{a})\cdot\left(  U_{2}t^{a}\right)  \cdot U_{3}%
\int_{-\infty}^{0}dz_{1}^{+}\int_{-\infty}^{0}dz_{2}^{+}G_{0}^{--}(z_{2}%
,z_{1})
\]%
\[
-g^{2}(t^{a}U_{1})\cdot(t^{a}U_{2})\cdot U_{3}\int_{0}^{\infty}dz_{2}^{+}%
\int_{0}^{\infty}dz_{1}^{+}G_{0}^{--}(z_{1},z_{2})
\]%
\[
=-g^{2}\left[  (U_{1}t^{a})\cdot\left(  U_{2}t^{a}\right)  \cdot U_{3}%
\int_{-\infty}^{0}dz_{1}^{+}\int_{-\infty}^{0}dz_{2}^{+}+(t^{a}U_{1}%
)\cdot(t^{a}U_{2})\cdot U_{3}\int_{0}^{\infty}dz_{2}^{+}\int_{0}^{\infty
}dz_{1}^{+}\right]
\]%
\begin{equation}
\times\int\frac{d^{4}p}{(2\pi)^{4}}e^{-ip^{-}z_{12}^{+}+i\,\vec{p}\,\vec
{z}_{12}}\frac{2ip^{-}}{p^{+}(p^{2}+i0)}.
\end{equation}
When we integrate (\ref{G--}) with respect to $p^{-}$ via residues, we see
that $p^{-}$ has a tiny positive imaginary part $p^{-}\rightarrow
p^{-}+i\varepsilon$ if $p^{+}<0$ and $x^{+}<0$ whereas $p^{-}\rightarrow
p^{-}-i\varepsilon$ if $p^{+}>0$ and $x^{+}>0.$ Therefore one can introduce
these imaginary parts into the integral for $B_{v}$ and take the integrals
with respect to $z_{1}^{+}$ and $z_{2}^{+}$ first. We have%
\[
B_{v12}=-g^{2}\int\frac{d^{4}p}{(2\pi)^{4}}\left[  (U_{1}t^{a})\cdot\left(
U_{2}t^{a}\right)  \cdot U_{3}\int_{-\infty}^{0}e^{-i(p^{-}+i\varepsilon
)z_{1}^{+}}dz_{1}^{+}\int_{-\infty}^{0}e^{i(p^{-}-i\varepsilon)z_{2}^{+}%
}dz_{2}^{+}\right.
\]%
\[
\left.  +(t^{a}U_{1})\cdot(t^{a}U_{2})\cdot U_{3}\int_{0}^{\infty}%
e^{i(p^{-}+i\varepsilon)z_{2}^{+}}dz_{2}^{+}\int_{0}^{\infty}e^{-i(p^{-}%
-i\varepsilon)z_{1}^{+}}dz_{1}^{+}\right]
\]%
\begin{equation}
=-g^{2}\int\frac{d^{4}p}{(2\pi)^{4}}\frac{2ip^{-}e^{i\,\vec{p}\,\vec{z}_{12}%
}\left[  (t^{a}U_{1})\cdot(t^{a}U_{2})+(U_{1}t^{a})\cdot\left(  U_{2}%
t^{a}\right)  \right]  \cdot U_{3}}{(p^{-}+i\varepsilon)(p^{-}-i\varepsilon
)p^{+}(2p^{+}p^{-}-\vec{p}^{\,\,2}+i0)}.
\end{equation}
Integrating $p^{-}$ out via residues one comes to
\[
B_{v12}=-g^{2}\left[  (t^{a}U_{1})\cdot(t^{a}U_{2})+(U_{1}t^{a})\cdot\left(
U_{2}t^{a}\right)  \right]  \cdot U_{3}%
\]%
\begin{equation}
\times\int\frac{d\vec{p}}{(2\pi)^{3}}\frac{e^{i\,\vec{p}\,\vec{z}_{12}}}%
{\vec{p}^{\,\,2}-i0}\int\left(  \theta\left(  p^{+}\right)  -\theta\left(
-p^{+}\right)  \right)  \frac{dp^{+}}{p^{+}}.
\end{equation}
Then, using
\begin{equation}
\int\frac{d\vec{p}}{(2\pi)^{3}}\frac{e^{i\,\vec{p}\,\vec{z}_{12}}}{\vec
{p}^{\,\,2}-i0}=\int\frac{d\vec{p}}{(2\pi)^{3}}e^{i\,\vec{p}\,\vec{z}_{12}%
}\int\frac{d\vec{z}_{4}d\vec{k}}{(2\pi)^{2}}e^{i(\vec{k}-\vec{p})\vec{z}_{42}%
}\frac{\vec{p}\vec{k}}{\vec{p}^{\,\,2}\vec{k}^{\,\,2}}=\int\frac{d\vec{z}_{4}%
}{(2\pi)^{3}}\frac{\left(  \vec{z}_{14}\vec{z}_{24}\right)  }{\vec{z}%
_{14}^{\,\,2}\vec{z}_{24}^{\,\,2}}%
\end{equation}
we get
\begin{equation}
B_{v12}=-\frac{\alpha_{s}}{\pi^{2}}\left[  (t^{a}U_{1})\cdot(t^{a}%
U_{2})+(U_{1}t^{a})\cdot\left(  U_{2}t^{a}\right)  \right]  \cdot U_{3}%
\int_{\sigma_{1}}^{\sigma}\frac{dp^{+}}{p^{+}}\int d\vec{z}_{4}\frac{\left(
\vec{z}_{14}\vec{z}_{24}\right)  }{\vec{z}_{14}^{\,\,2}\vec{z}_{24}^{\,\,2}}.
\label{Bv}%
\end{equation}
Quite similarly, the virtual contribution from the interaction of $U_{1}$
reads
\[
B_{v1}=-g^{2}(U_{1}t^{a}t^{a})\cdot U_{2}\cdot U_{3}\int_{-\infty}^{0}%
dz_{1}^{+}\int_{-\infty}^{0}dz_{1}^{\prime+}\frac{1}{2}G_{0}^{--}(z_{1}%
,z_{1}^{\prime})
\]%
\[
-g^{2}(t^{a}t^{a}U_{1})\cdot U_{2}\cdot U_{3}\int_{0}^{\infty}dz_{1}^{+}%
\int_{0}^{\infty}dz_{1}^{\prime+}\frac{1}{2}G_{0}^{--}(z_{1},z_{1}^{\prime})
\]%
\begin{equation}
=-\frac{\alpha_{s}}{2\pi^{2}}\left[  (t^{a}t^{a}U_{1})\cdot U_{2}+(U_{1}%
t^{a}t^{a})\cdot U_{2}\right]  \cdot U_{3}\int_{\sigma_{1}}^{\sigma}%
\frac{dp^{+}}{p^{+}}\int\frac{d\vec{z}_{4}}{\vec{z}_{14}^{\,\,2}}.
\end{equation}
Collecting all the contributions and differentiating with respect to $\eta
=\ln\sigma,$ one gets the equation%
\begin{equation}
\frac{\partial}{\partial\eta}\langle B_{123}^{\eta}\rangle=\langle K\otimes
B_{123}^{\eta}\rangle. \label{d/detaU=KU}%
\end{equation}
Its explicit form reads%
\[
\frac{\partial}{\partial\eta}\langle B_{123}^{\eta}\rangle=\frac{\alpha_{s}%
}{\pi^{2}}\int d\vec{z}_{4}\left(  \left[  \frac{\left(  \vec{z}_{41}\vec
{z}_{42}\right)  }{\vec{z}_{41}^{\,\,2}\vec{z}_{42}^{\,\,2}}U_{3}\cdot\left\{
U_{4}^{ba}\left(  (t^{b}U_{1})\cdot(U_{2}t^{a})+(U_{1}t^{a})\cdot(t^{b}%
U_{2})\right)  \right.  \right.  \right.
\]%
\[
\left.  -\left(  (t^{a}U_{1})\cdot(t^{a}U_{2})+(U_{1}t^{a})\cdot\left(
U_{2}t^{a}\right)  \right)  \right\}  \left.  +(1\leftrightarrow
3)+(2\leftrightarrow3)\frac{{}}{{}}\right]
\]%
\begin{equation}
+\left.  \left[  \frac{1}{\vec{z}_{41}^{\,\,2}}\left\{  U_{4}^{ba}\left(
t^{b}U_{1}t^{a}\right)  -\frac{1}{2}\left(  t^{a}t^{a}U_{1}\right)  -\frac
{1}{2}\left(  U_{1}t^{a}t^{a}\right)  \right\}  \cdot U_{2}\cdot
U_{3}+(1\leftrightarrow2)+(1\leftrightarrow3)\right]  \right)  .
\label{evolutionEq1}%
\end{equation}
Then we can use the SU(3) identity%
\begin{equation}
\left(  U_{2}U_{4}^{\dag}U_{1}+U_{1}U_{4}^{\dag}U_{2}\right)  \cdot U_{4}\cdot
U_{3}=-B_{123}^{\eta}+\frac{1}{2}(B_{144}^{\eta}B_{324}^{\eta}+B_{244}^{\eta
}B_{314}^{\eta}-B_{344}^{\eta}B_{214}^{\eta}), \label{IDENTITY}%
\end{equation}
and
\begin{equation}
U_{4}^{ba}=2tr(t^{b}U_{4}t^{a}U_{4}^{\dag}),\quad t_{ij}^{a}t_{kl}^{a}%
=\frac{1}{2}\delta_{il}\delta_{kj}-\frac{1}{2N_{c}}\delta_{ij}\delta_{kl}
\label{Uadjoint}%
\end{equation}
to rearrange this expression in the following way
\[
\frac{\partial B_{123}^{\eta}}{\partial\eta}=\frac{\alpha_{s}3}{4\pi^{2}}\int
d\vec{z}_{4}\left[  \frac{\vec{z}_{12}^{\,\,2}}{\vec{z}_{41}^{\,\,2}\vec
{z}_{42}^{\,\,2}}(-B_{123}^{\eta}+\frac{1}{6}(B_{144}^{\eta}B_{324}^{\eta
}+B_{244}^{\eta}B_{314}^{\eta}-B_{344}^{\eta}B_{214}^{\eta}))\right.
\]%
\begin{equation}
\left.  \frac{{}}{{}}+(1\leftrightarrow3)+(2\leftrightarrow3)\right]  ,
\label{evolutionEq2}%
\end{equation}
where we dropped the angular brackets for brevity.

\section{Diagrams proportional to the LO$^{2}$ kernel}%
\begin{figure}[ptbh]
\begin{center}
\includegraphics[
width=\textwidth
]{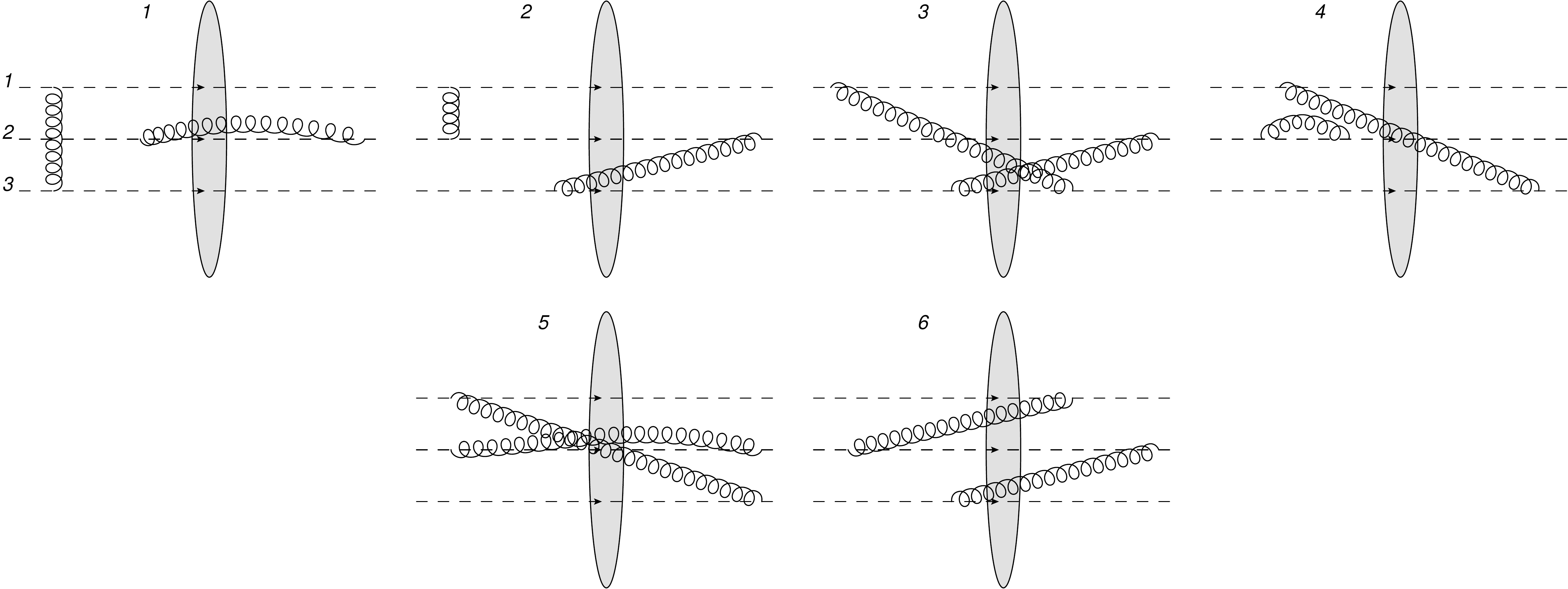}
\end{center}
\caption{Diagrams contributing to LO$^{2}$ kernel. The dashed lines represent
the Wilson lines with $z^{-}=0$ and $\vec{z}_{1,2,3}$ along the $z^{+}$ axis
from $z^{+}=-\infty$ to the left to $z^{+}=+\infty$ to the right. The grey
ellipse stands for the shockwave at $z^{+}=0.$ }%
\label{LO2fig}%
\end{figure}

The connected part of the NLO kernel comes from the diagrams where all the
three Wilson lines have nontrivial evolution. The first group of such diagrams
is depicted in fig \ref{LO2fig}. These diagrams and the diagrams which they
come into after the reflection with respect to the shock wave and after all
possible permutations of $U_{1},U_{2},U_{3}$ can be totally reduced to the
second iteration of the LO kernel. Indeed, the first diagram reads%
\[
\langle B_{123}^{\eta}\rangle|_{1}=g^{4}(U_{1}t^{a})\cdot(t^{b^{\prime}}%
U_{2}t^{b})\cdot(U_{3}t^{a})
\]%
\[
\times\int_{-\infty}^{0}dz_{1}^{+}\int_{-\infty}^{0}dz_{3}^{+}G_{0}^{--}%
(z_{3},z_{1})\int_{-\infty}^{0}dz_{2}^{\prime+}\int_{0}^{\infty}dz_{2}%
^{+}G^{--}(z_{2},z_{2}^{\prime})^{b^{\prime}b}%
\]%
\begin{equation}
=-\frac{\alpha_{s}^{2}}{\pi^{4}}(U_{1}t^{a})\cdot(t^{b^{\prime}}U_{2}%
t^{b})\cdot(U_{3}t^{a})\int_{\sigma_{1}}^{\sigma}\frac{dp^{+}}{p^{+}}%
\int_{\sigma_{1}}^{\sigma}\frac{dk^{+}}{k^{+}}\int\frac{d\vec{z}_{4}}{\vec
{z}_{42}^{\,\,2}}U_{4}^{b^{\prime}b}\int d\vec{z}_{0}\frac{\left(  \vec
{z}_{01}\vec{z}_{03}\right)  }{\vec{z}_{01}^{\,\,2}\vec{z}_{03}^{\,\,2}}.
\end{equation}
Hence%
\begin{equation}
\frac{\partial}{\partial\eta}\langle B_{123}^{\eta}\rangle|_{1}=-2\ln
\frac{\sigma}{\sigma_{1}}\frac{\alpha_{s}^{2}}{\pi^{4}}(U_{1}t^{a}%
)\cdot(t^{b^{\prime}}U_{2}t^{b})\cdot(U_{3}t^{a})\int\frac{d\vec{z}_{4}}%
{\vec{z}_{42}^{\,\,2}}U_{4}^{b^{\prime}b}\int d\vec{z}_{0}\frac{\left(
\vec{z}_{01}\vec{z}_{03}\right)  }{\vec{z}_{01}^{\,\,2}\vec{z}_{03}^{\,\,2}}.
\end{equation}
Here we used expressions (\ref{Br1}) and (\ref{Bv}) from the previous section.
In the NLO equation (\ref{d/detaU=KU}) changes into%
\begin{equation}
\frac{\partial}{\partial\eta}\langle B_{123}^{\eta}\rangle=\langle
K_{LO}\otimes B_{123}^{\eta}\rangle+\langle K_{NLO}\otimes B_{123}^{\eta
}\rangle.
\end{equation}
Therefore
\begin{equation}
\frac{\partial}{\partial\eta}\langle B_{123}^{\eta}\rangle-\langle
K_{LO}\otimes B_{123}^{\eta}\rangle=\langle K_{NLO}\otimes B_{123}^{\eta
}\rangle.
\end{equation}
One can obtain $\langle K_{LO}\otimes B_{123}^{\eta}\rangle$ applying the LO
evolution to the Wilson lines in the r.h.s. of LO evolution equation
(\ref{evolutionEq1}). Among others, it contains these 2 terms
\begin{equation}
\frac{\alpha_{s}}{\pi^{2}}\int d\vec{z}_{4}\left[  \frac{1}{\vec{z}%
_{42}^{\,\,2}}U_{4}^{ba}U_{1}\cdot(t^{b}U_{2}t^{a})\cdot U_{3}-\frac{(\vec
{z}_{41}\vec{z}_{43})}{\vec{z}_{41}^{\,\,2}\vec{z}_{43}^{\,\,2}}(U_{1}%
t^{a})\cdot U_{2}\cdot(U_{3}t^{a})+\dots\right]  .
\end{equation}
If we calculate these terms in the shock wave background in the LO, we will
have among others the contribution where we dress the Wilson lines $U_{1}$ and
$U_{3}$ from the first term and $U_{2}$ from the second term%
\[
\langle K_{LO}\otimes B_{123}^{\eta}\rangle|_{1}=\frac{\alpha_{s}^{2}}{\pi
^{4}}\int_{\sigma_{1}}^{\sigma}\frac{dp^{+}}{p^{+}}\int d\vec{z}_{0}\int
d\vec{z}_{4}%
\]%
\[
\times\left[  -\frac{1}{\vec{z}_{42}^{\,\,2}}\frac{(\vec{z}_{01}\vec{z}_{03}%
)}{\vec{z}_{01}^{\,\,2}\vec{z}_{03}^{\,\,2}}U_{4}^{ba}(U_{1}t^{c})\cdot(
t^{b}U_{2}t^{a}) \cdot(U_{3}t^{c})-\frac{(\vec{z}_{41}\vec{z}_{43})}{\vec
{z}_{41}^{\,\,2}\vec{z}_{43}^{\,\,2}}\frac{1}{\vec{z}_{02}^{\,\,2}}(U_{1}%
t^{a})\cdot(U_{0}^{dc}t^{d}U_{2}t^{c})\cdot(U_{3}t^{a})\right]
\]%
\begin{equation}
=\frac{\partial}{\partial\eta}\langle B_{123}^{\eta}\rangle|_{1}.
\end{equation}
As a result, this diagram does not contribute to the NLO kernel. The same is
true for all the diagrams in fig. \ref{LO2fig}.

\section{Diagrams with 2 gluons intersecting the shockwave}%
\begin{figure}
[ptbh]
\begin{center}
\includegraphics[
width= \textwidth
]%
{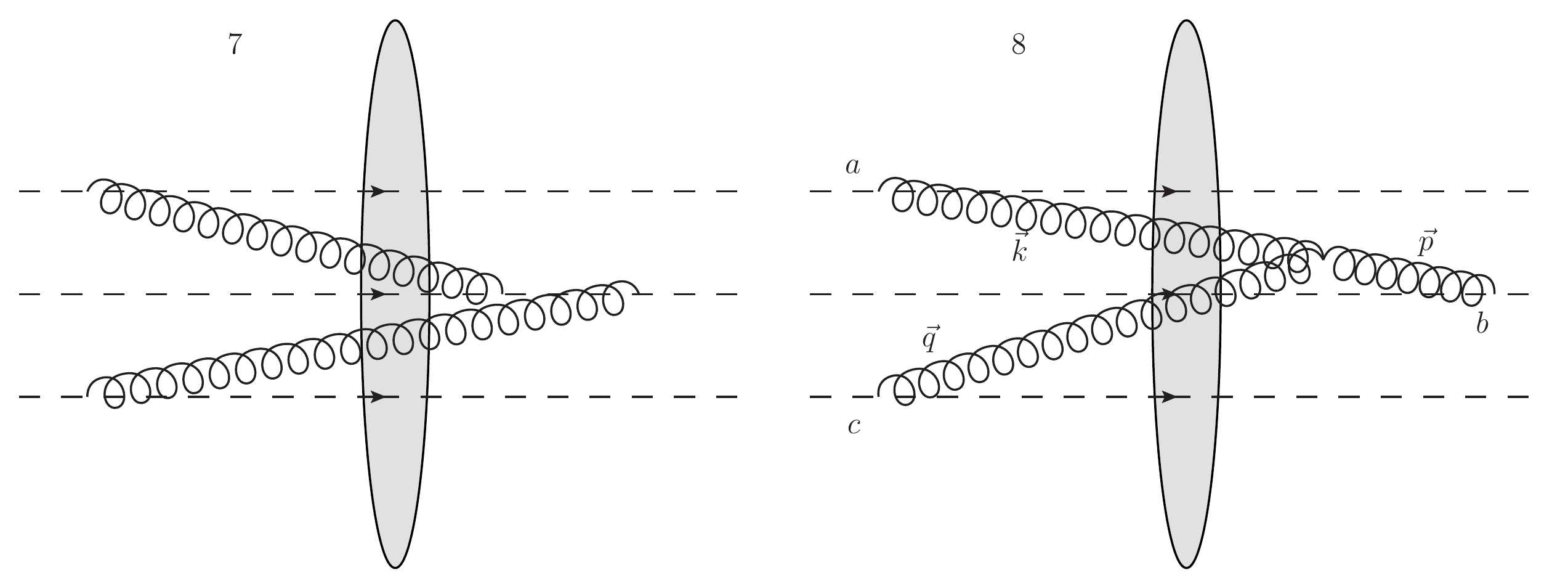}%
\caption{Diagrams with two gluons intersecting the shockwave.}%
\label{2gluonDiags}%
\end{center}
\end{figure}
Next we consider the diagrams with two gluons intersecting the shockwave
depicted in fig \ref{2gluonDiags}. Diagram 7 reads (here $z_{2}^{\prime
}=(z_{2}^{\prime+},0,\vec{z}_{2})$)%
\[
\langle B_{123}^{\eta}\rangle|_{7}=g^{4}(U_{1}t^{a})\cdot(t^{b^{\prime}%
}t^{a^{\prime}}U_{2})\cdot(U_{3}t^{b})
\]%
\[
\times\int_{-\infty}^{0}dz_{1}^{+}\int_{-\infty}^{0}dz_{3}^{+}\int_{0}%
^{\infty}dz_{2}^{+}\int_{z_{2}^{+}}^{\infty}dz_{2}^{\prime+}G^{--}(z_{2}%
,z_{1})^{a^{\prime}a}G^{--}(z_{2}^{\prime},z_{3})^{b^{\prime}b}%
\]%
\[
=g^{4}(U_{1}t^{a})\cdot(t^{b^{\prime}}t^{a^{\prime}}U_{2})\cdot(U_{3}%
t^{b})\int_{\sigma_{1}}^{\sigma}\frac{2p^{+}dp^{+}}{\left(  2\pi\right)  ^{3}%
}\int_{\sigma_{1}}^{\sigma}\frac{2k^{+}dk^{+}}{\left(  2\pi\right)  ^{3}}%
\int\left(  \vec{z}_{20}\vec{z}_{10}\right)  U_{0}^{a^{\prime}a}d\vec{z}_{0}%
\]%
\[
\times\int\left(  \vec{z}_{24}\vec{z}_{34}\right)  U_{4}^{b^{\prime}b}d\vec
{z}_{4}\int_{-\infty}^{0}\frac{dz_{1}^{+}}{2(z_{1}^{+})^{2}}e^{-ip^{+}%
\frac{\vec{z}_{01}{}^{2}+i0}{2z_{1}^{+}}}\int_{-\infty}^{0}\frac{dz_{3}^{+}%
}{2(z_{3}^{+})^{2}}e^{-ik^{+}\frac{\vec{z}_{34}{}^{2}+i0}{2z_{3}^{+}}}%
\]%
\begin{equation}
\times\int_{0}^{\infty}\frac{dz_{2}^{\prime+}}{2(z_{2}^{\prime+})^{2}%
}e^{ik^{+}\frac{\vec{z}_{24}{}^{2}+i0}{2z_{2}^{\prime+}}}\int_{0}%
^{z_{2}^{\prime+}}\frac{dz_{2}^{+}}{2(z_{2}^{+})^{2}}e^{ip^{+}\frac{\vec
{z}_{02}{}^{2}+i0}{2z_{2}^{+}}}.
\end{equation}%
\[
\langle B_{123}^{\eta}\rangle|_{7}=4g^{4}(U_{1}t^{a})\cdot(t^{b^{\prime}%
}t^{a^{\prime}}U_{2})\cdot(U_{3}t^{b})\int\frac{\left(  \vec{z}_{20}\vec
{z}_{10}\right)  }{\vec{z}_{02}{}^{2}\vec{z}_{01}{}^{2}}U_{0}^{a^{\prime}%
a}\frac{d\vec{z}_{0}}{\left(  2\pi\right)  ^{3}}\int\frac{\left(  \vec{z}%
_{24}\vec{z}_{34}\right)  }{\vec{z}_{34}{}^{2}}U_{4}^{b^{\prime}b}\frac
{d\vec{z}_{4}}{\left(  2\pi\right)  ^{3}}%
\]%
\begin{equation}
\times\int_{\sigma_{1}}^{\sigma}\frac{dp^{+}}{p^{+}}\int_{\sigma_{1}}^{\sigma
}\frac{dk^{+}}{k^{+}\vec{z}_{24}{}^{2}+\vec{z}_{02}{}^{2}p^{+}}.
\end{equation}
The corresponding term in $\langle K_{LO}\otimes B_{123}^{\eta}\rangle$ comes
from the following term in LO evolution equation (\ref{evolutionEq1})
\begin{equation}
\frac{\alpha_{s}}{\pi^{2}}\int d\vec{z}_{4}\frac{(\vec{z}_{43}\vec{z}_{42}%
)}{\vec{z}_{43}^{\,\,2}\vec{z}_{42}^{\,\,2}}U_{4}^{ba}(U_{3}t^{a})\cdot
(t^{b}U_{2})\cdot U_{1}.
\end{equation}
If we dress $U_{1}$ and $U_{2}$ in this expression, we get one of the
contributions which acts as a subtraction term for diagram 7%
\begin{equation}
\langle K_{LO}\otimes B_{123}^{\eta}\rangle|_{7}=\frac{\alpha_{s}^{2}}{\pi
^{4}}\int_{\sigma_{1}}^{\sigma}\frac{dp^{+}}{p^{+}}\int d\vec{z}_{0}U_{0}%
^{cd}\frac{\left(  \vec{z}_{01}\vec{z}_{02}\right)  }{\vec{z}_{01}^{\,\,2}%
\vec{z}_{02}^{\,\,2}}\int d\vec{z}_{4}U_{4}^{ba}\frac{\left(  \vec{z}_{43}%
\vec{z}_{42}\right)  }{\vec{z}_{43}^{\,\,2}\vec{z}_{42}^{\,\,2}}(U_{1}%
t^{d})\cdot(t^{b}t^{c}U_{2})\cdot(U_{3}t^{a}).
\end{equation}
Then
\[
\langle K_{NLO}\otimes B_{123}^{\eta}\rangle|_{7}=\frac{\partial}{\partial
\eta}\langle B_{123}^{\eta}\rangle|_{7}-\langle K_{LO}\otimes B_{123}^{\eta
}\rangle|_{7}%
\]%
\begin{equation}
=\frac{\alpha_{s}^{2}}{\pi^{4}}(U_{1}t^{a})\cdot(t^{b^{\prime}}t^{a^{\prime}%
}U_{2})\cdot(U_{3}t^{b})\int d\vec{z}_{0}\frac{\left(  \vec{z}_{20}\vec
{z}_{10}\right)  }{\vec{z}_{02}{}^{2}\vec{z}_{01}{}^{2}}U_{0}^{a^{\prime}%
a}\int d\vec{z}_{4}\frac{\left(  \vec{z}_{24}\vec{z}_{34}\right)  }{\vec
{z}_{42}^{\,\,2}\vec{z}_{34}{}^{2}}U_{4}^{b^{\prime}b}\ln\frac{\vec{z}%
_{42}^{\,\,2}}{\vec{z}_{02}^{\,\,2}{}}.
\end{equation}
The contribution of diagram 7 with the interchange of 1st and 3rd Wilson lines
can be obtained via the interchange $1\leftrightarrow3$ in this result. For
the sum of these diagrams one gets
\[
\langle K_{NLO}\otimes B_{123}^{\eta}\rangle|_{7+7(1\leftrightarrow3)}
\]%
\begin{equation}
=\frac{\alpha_{s}^{2}}{\pi^{4}}if^{a^{\prime}bc^{\prime}}(U_{1}t^{a}%
)\cdot(t^{b}U_{2})\cdot(U_{3}t^{c})\int d\vec{z}_{0}\frac{\left(  \vec{z}%
_{20}\vec{z}_{10}\right)  }{\vec{z}_{02}{}^{2}\vec{z}_{01}{}^{2}}%
U_{0}^{a^{\prime}a}\int d\vec{z}_{4}\frac{\left(  \vec{z}_{24}\vec{z}%
_{34}\right)  }{\vec{z}_{24}^{\,\,2}\vec{z}_{34}{}^{2}}U_{4}^{c^{\prime}c}%
\ln\frac{\vec{z}_{42}^{\,\,2}}{\vec{z}_{02}^{\,\,2}{}}.
\end{equation}
The contribution of the diagram which is a mirror reflection of diagram 7 with
respect to the shockwave reads%
\[
\langle B_{123}^{\eta}\rangle|_{7m}=g^{4}(t^{a^{\prime}}U_{1})\cdot(U_{2}%
t^{a}t^{b})\cdot(t^{b^{\prime}}U_{3})
\]%
\begin{equation}
\times\int_{0}^{\infty}dz_{1}^{+}\int_{0}^{\infty}dz_{3}^{+}\int_{-\infty}%
^{0}dz_{2}^{+}\int_{-\infty}^{z_{2}^{+}}dz_{2}^{\prime+}G^{--}(z_{1}%
,z_{2})^{a^{\prime}a}G^{--}(z_{3},z_{2}^{\prime})^{b^{\prime}b}.
\end{equation}
Its integrated form differs from the contribution of diagram 7 only in the $t$
matrix order. One gets%
\begin{equation}
\langle K_{NLO}\otimes B_{123}^{\eta}\rangle|_{7m}=\frac{\alpha_{s}^{2}}%
{\pi^{4}}(t^{a^{\prime}}U_{1})\cdot(U_{2}t^{a}t^{b})\cdot(t^{b^{\prime}}%
U_{3})\int d\vec{z}_{0}\frac{\left(  \vec{z}_{20}\vec{z}_{10}\right)  }%
{\vec{z}_{02}{}^{2}\vec{z}_{01}{}^{2}}U_{0}^{a^{\prime}a}\int d\vec{z}%
_{4}\frac{\left(  \vec{z}_{24}\vec{z}_{34}\right)  }{\vec{z}_{42}^{\,\,2}%
\vec{z}_{34}{}^{2}}U_{4}^{b^{\prime}b}\ln\frac{\vec{z}_{42}^{\,\,2}}{\vec
{z}_{02}^{\,\,2}{}}.
\end{equation}
The corresponding diagram with the interchange of 1st and 3rd Wilson lines
appears from this expression after the substitution $(U_{2}t^{a}%
t^{b})\rightarrow-(U_{2}t^{b}t^{a}).$ The sum of these diagrams is%
\[
\langle K_{NLO}\otimes B_{123}^{\eta}\rangle|_{7m+7m(1\leftrightarrow3)}
\]%
\begin{equation}
=\frac{\alpha_{s}^{2}}{\pi^{4}}if^{abc}(t^{a^{\prime}}U_{1})\cdot(U_{2}%
t^{c})\cdot(t^{b^{\prime}}U_{3})\int d\vec{z}_{0}\frac{\left(  \vec{z}%
_{20}\vec{z}_{10}\right)  }{\vec{z}_{02}{}^{2}\vec{z}_{01}{}^{2}}%
U_{0}^{a^{\prime}a}\int d\vec{z}_{4}\frac{\left(  \vec{z}_{24}\vec{z}%
_{34}\right)  }{\vec{z}_{42}^{\,\,2}\vec{z}_{34}{}^{2}}U_{4}^{b^{\prime}b}%
\ln\frac{\vec{z}_{42}^{\,\,2}}{\vec{z}_{02}^{\,\,2}{}}.
\end{equation}
Finally, the contribution of the four diagrams: 7, 7($1\leftrightarrow3$) and
their mirror reflections with respect to the shockwave has the form%
\[
\langle K_{NLO}\otimes B_{123}^{\eta}\rangle|_{7+7(1\leftrightarrow
3)+7m+7(1\leftrightarrow3)m}=\frac{\alpha_{s}^{2}}{\pi^{4}}\int d\vec{z}%
_{0}\frac{\left(  \vec{z}_{20}\vec{z}_{10}\right)  }{\vec{z}_{02}{}^{2}\vec
{z}_{01}{}^{2}}U_{0}^{a^{\prime}a}\int d\vec{z}_{4}\frac{\left(  \vec{z}%
_{24}\vec{z}_{34}\right)  }{\vec{z}_{24}^{\,\,2}\vec{z}_{34}{}^{2}}%
U_{4}^{c^{\prime}c}\ln\frac{\vec{z}_{42}^{\,\,2}}{\vec{z}_{02}^{\,\,2}{}}%
\]%
\begin{equation}
\times i\left\{  f^{a^{\prime}bc^{\prime}}(U_{1}t^{a})\cdot(t^{b}U_{2}%
)\cdot(U_{3}t^{c})-f^{abc}(t^{a^{\prime}}U_{1})\cdot(U_{2}t^{b})\cdot
(t^{c^{\prime}}U_{3})\right\}  . \label{diagram7+all}%
\end{equation}
Let us turn to diagram 8. It reads%
\[
\langle B_{123}^{\eta}\rangle|_{8}=-g^{4}f^{a^{\prime}b^{\prime}c^{\prime}%
}(U_{1}t^{a})\cdot(t^{b}U_{2})\cdot(U_{3}t^{c})\int_{-\infty}^{0}dz_{1}%
^{+}\int_{-\infty}^{0}dz_{3}^{+}\int_{0}^{\infty}dz_{2}^{+}\int\theta
(x^{+})d^{4}x
\]%
\[
\times\left\{  \frac{\partial G^{a^{\prime}a}(x,z_{1})_{j}^{\,\,\,\,\,-}%
}{\partial x^{\mu}}\left[  G_{0}^{bb^{\prime}}(z_{2},x)^{-\mu}G^{c^{\prime}%
c}(x,z_{3})^{j-}-G_{0}^{bb^{\prime}}(z_{2},x)^{-j}G^{c^{\prime}c}%
(x,z_{3})^{\mu-}\right]  \right.
\]%
\[
+\frac{\partial G_{0}^{bb^{\prime}}(z_{2},x)_{\,\,\,j}^{-}}{\partial x^{\mu}%
}\left[  G^{a^{\prime}a}(x,z_{1})^{j-}G^{c^{\prime}c}(x,z_{3})^{\mu
-}-G^{a^{\prime}a}(x,z_{1})^{\mu-}G^{c^{\prime}c}(x,z_{3})^{j-}\right]
\]%
\begin{equation}
\left.  +\frac{\partial G^{c^{\prime}c}(x,z_{3})_{j}^{\,\,\,\,\,-}}{\partial
x^{\mu}}\left[  G^{a^{\prime}a}(x,z_{1})^{\mu-}G_{0}^{bb^{\prime}}%
(z_{2},x)^{-j}-G^{a^{\prime}a}(x,z_{1})^{j-}G_{0}^{bb^{\prime}}(z_{2}%
,x)^{-\mu}\right]  \right\}  .
\end{equation}
Here we sum over $j=1,2$ and $\mu=-,1,2.$ It is convenient to split this
expression into two parts.%
\begin{equation}
\langle B_{123}^{\eta}\rangle|_{8}=\langle B_{123}^{\eta}\rangle|_{8_{1}%
}+\langle B_{123}^{\eta}\rangle|_{8_{2}}.
\end{equation}%
\[
\langle B_{123}^{\eta}\rangle|_{8_{1}}=-g^{4}f^{a^{\prime}b^{\prime}c^{\prime
}}(U_{1}t^{a})\cdot(t^{b}U_{2})\cdot(U_{3}t^{c})\int_{-\infty}^{0}dz_{1}%
^{+}\int_{-\infty}^{0}dz_{3}^{+}\int_{0}^{\infty}dz_{2}^{+}\int\theta
(x^{+})d^{4}x
\]%
\begin{equation}
\times G_{0}^{bb^{\prime}}(z_{2},x)^{--}\left\{  G^{c^{\prime}c}(x,z_{3}%
)^{j-}\frac{\partial G^{a^{\prime}a}(x,z_{1})_{j}^{\,\,\,\,\,-}}{\partial
x^{-}}-G^{a^{\prime}a}(x,z_{1})^{j-}\frac{\partial G^{c^{\prime}c}%
(x,z_{3})_{j}^{\,\,\,\,\,-}}{\partial x^{-}}\right\}  .
\end{equation}%
\[
\langle B_{123}^{\eta}\rangle|_{8_{1}}=-g^{4}f^{a^{\prime}b^{\prime}c^{\prime
}}(U_{1}t^{a})\cdot(t^{b}U_{2})\cdot(U_{3}t^{c})\int_{-\infty}^{0}dz_{1}%
^{+}\int_{-\infty}^{0}dz_{3}^{+}\int_{0}^{\infty}dz_{2}^{+}\int\theta
(x^{+})d^{4}x
\]%
\begin{equation}
\times\int_{\sigma_{1}}^{\sigma}\frac{dp^{+}}{2\pi}e^{ix^{-}p^{+}}\int
\frac{d\vec{p}dp^{-}}{(2\pi)^{3}}\frac{2ip^{-}e^{-i(p^{-}-i\varepsilon
)z_{2}^{+}+ip^{-}x^{+}+i\,\vec{p}\,\vec{z}_{2x}}}{p^{+}(p^{2}+i0)}\left\{
\dots\frac{{}}{{}}\right\}  =0.
\end{equation}
Here as in the LO calculation, we changed
\begin{equation}
e^{-ip^{-}z_{2x}^{+}}\rightarrow e^{-i(p^{-}-i\varepsilon)z_{2}^{+}%
+ip^{-}x^{+}}%
\end{equation}
and used (\ref{G--integral1=0}) to get zero. Therefore%
\[
\langle B_{123}^{\eta}\rangle|_{8}=\langle B_{123}^{\eta}\rangle|_{8_{2}%
}=g^{4}f^{a^{\prime}bc^{\prime}}(U_{1}t^{a})\cdot(t^{b}U_{2})\cdot(U_{3}%
t^{c})
\]%
\[
\times\int_{0}^{\infty}\frac{dz_{2}^{+}}{2(z_{2x}^{+})^{2}}\int\frac{d^{4}%
x}{(x^{+})^{3}}\theta(z_{2x}^{+})\int U_{0}^{a^{\prime}a}\frac{d\vec{z}_{0}%
}{\vec{z}_{10}^{\,\,2}}\int U_{4}^{c^{\prime}c}\frac{d\vec{z}_{4}}{\vec
{z}_{34}^{\,\,2}}%
\]%
\[
\times\int_{\sigma_{1}}^{\sigma}\frac{dp^{+}}{(2\pi)^{2}}e^{ip^{+}\frac
{\vec{z}_{2x}^{\,\,2}+i0}{2z_{2x}^{+}}}\int_{\sigma_{1}}^{\sigma-\sigma_{1}%
}\frac{dk^{+}}{\left(  2\pi\right)  ^{3}}e^{ik^{+}\frac{\vec{z}_{x0}%
^{\,\,2}+i0}{2x^{+}}}\int_{\sigma_{1}}^{\sigma-\sigma_{1}}\frac{dq^{+}%
}{\left(  2\pi\right)  ^{3}}e^{iq^{+}\frac{\vec{z}_{x4}^{\,\,2}+i0}{2x^{+}}%
}e^{ix^{-}\left[  p^{+}-q^{+}-k^{+}\right]  }%
\]%
\[
\times\left[  k^{+}(\vec{z}_{10}\vec{z}_{2x})(\vec{z}_{34}\vec{z}_{4x}%
)+p^{+}\left[  (\vec{z}_{10}\vec{z}_{2x})(\vec{z}_{34}\vec{z}_{4x})-(\vec
{z}_{2x}\vec{z}_{34})\left(  \vec{z}_{10}\vec{z}_{0x}\right)  \right]
-q^{+}\left(  \vec{z}_{10}\vec{z}_{0x}\right)  (\vec{z}_{2x}\vec{z}_{34}%
)\frac{{}}{{}}\right.
\]%
\begin{equation}
\left.  +\frac{{}}{{}}k^{+}\left[  -\left(  \vec{z}_{10}\vec{z}_{34}\right)
(\vec{z}_{2x}\vec{z}_{x0})+\left(  \vec{z}_{10}\vec{z}_{2x}\right)  (\vec
{z}_{x0}\vec{z}_{34})\right]  +q^{+}\left[  -(\vec{z}_{2x}\vec{z}_{34}%
)(\vec{z}_{10}\vec{z}_{x4})+(\vec{z}_{2x}\vec{z}_{x4})\left(  \vec{z}_{10}%
\vec{z}_{34}\right)  \right]  \right]  .
\end{equation}
Then integrating with respect to $z_{2x}^{+}$ and $x^{+}$ we arrive to%
\[
\langle B_{123}^{\eta}\rangle|_{8}=-4g^{4}if^{a^{\prime}bc^{\prime}}%
(U_{1}t^{a})\cdot(t^{b}U_{2})\cdot(U_{3}t^{c})\int U_{0}^{a^{\prime}a}%
\frac{d\vec{z}_{0}}{\vec{z}_{10}^{\,\,2}}\int U_{4}^{c^{\prime}c}\frac
{d\vec{z}_{4}}{\vec{z}_{34}^{\,\,2}}%
\]%
\[
\int_{\sigma_{1}}^{\sigma}\frac{dp^{+}}{p^{+}(2\pi)^{2}}\int_{\sigma_{1}%
}^{\sigma-\sigma_{1}}\frac{dk^{+}}{\left(  2\pi\right)  ^{3}}\int_{\sigma_{1}%
}^{\sigma-\sigma_{1}}\frac{dq^{+}}{\left(  2\pi\right)  ^{3}}\int d\vec
{x}dx^{-}e^{ix^{-}\left[  p^{+}-q^{+}-k^{+}\right]  }\frac{1}{\vec{z}%
_{2x}^{\,\,2}}\frac{1}{(q^{+}\vec{z}_{x4}^{\,\,2}+k^{+}\vec{z}_{x0}%
^{\,\,2})^{2}}%
\]%
\[
\times\left[  p^{+}\left[  (\vec{z}_{10}\vec{z}_{20})(\vec{z}_{34}\vec{z}%
_{4x})-(\vec{z}_{24}\vec{z}_{34})\left(  \vec{z}_{10}\vec{z}_{0x}\right)
\right]  \frac{{}}{{}}\right.
\]%
\begin{equation}
\left.  +\frac{{}}{{}}k^{+}\left[  -\left(  \vec{z}_{10}\vec{z}_{34}\right)
(\vec{z}_{2x}\vec{z}_{x0})+\left(  \vec{z}_{10}\vec{z}_{2x}\right)  (\vec
{z}_{40}\vec{z}_{34})\right]  +q^{+}\left[  -(\vec{z}_{2x}\vec{z}_{34}%
)(\vec{z}_{10}\vec{z}_{04})+(\vec{z}_{2x}\vec{z}_{x4})\left(  \vec{z}_{10}%
\vec{z}_{34}\right)  \right]  \right]  .
\end{equation}
The integral with respect to $\vec{x}$ can be calculated by the Feynman
parameter technique. As a result%
\[
\langle B_{123}^{\eta}\rangle|_{8}=-2g^{4}if^{a^{\prime}bc^{\prime}}%
(U_{1}t^{a})\cdot(t^{b}U_{2})\cdot(U_{3}t^{c})\int U_{0}^{a^{\prime}a}%
\frac{d\vec{z}_{0}}{\vec{z}_{10}^{\,\,2}}\int U_{4}^{c^{\prime}c}\frac
{d\vec{z}_{4}}{\vec{z}_{34}^{\,\,2}}\int_{\sigma_{1}}^{\sigma}\frac{dp^{+}%
}{\left(  2\pi\right)  ^{3}}\int_{\sigma_{1}}^{p^{+}-\sigma_{1}}\frac{dk^{+}%
}{\left(  2\pi\right)  ^{3}}%
\]%
\begin{equation}
\times\frac{2}{\vec{z}_{04}^{\,\,2}((p^{+}-k^{+})\,\vec{z}_{42}^{\,\,2}%
+k^{+}\,\vec{z}_{02}^{\,\,2})}\left[  \frac{\left(  \vec{z}_{10}\vec{z}%
_{34}\right)  \left[  \vec{z}_{24}^{\,\,2}-\vec{z}_{02}^{\,\,2}\right]
}{2p^{+}\,}+\frac{(\vec{z}_{10}\vec{z}_{40})(\vec{z}_{24}\vec{z}_{34})}%
{k^{+}\,}-\frac{(\vec{z}_{04}\vec{z}_{34})(\vec{z}_{10}\vec{z}_{20})}%
{p^{+}-k^{+}}\right]  .
\end{equation}
Therefore%
\[
\frac{\partial}{\partial\eta}\langle B_{123}^{\eta}\rangle|_{8}=-2g^{4}%
if^{a^{\prime}bc^{\prime}}(U_{1}t^{a})\cdot(t^{b}U_{2})\cdot(U_{3}t^{c})\int
U_{0}^{a^{\prime}a}\frac{d\vec{z}_{0}}{\vec{z}_{10}^{\,\,2}}\int
U_{4}^{c^{\prime}c}\frac{d\vec{z}_{4}}{\vec{z}_{34}^{\,\,2}}\int_{\sigma_{1}%
}^{\sigma-\sigma_{1}}\frac{dk^{+}}{\left(  2\pi\right)  ^{6}}%
\]%
\begin{equation}
\times\frac{2\sigma}{\vec{z}_{04}^{\,\,2}((\sigma-k^{+})\,\vec{z}_{42}%
^{\,\,2}+k^{+}\,\vec{z}_{02}^{\,\,2})}\left[  \frac{\left(  \vec{z}_{10}%
\vec{z}_{34}\right)  \left[  \vec{z}_{24}^{\,\,2}-\vec{z}_{02}^{\,\,2}\right]
}{2\sigma}+\frac{(\vec{z}_{10}\vec{z}_{40})(\vec{z}_{24}\vec{z}_{34})}%
{k^{+}\,}-\frac{(\vec{z}_{04}\vec{z}_{34})(\vec{z}_{10}\vec{z}_{20})}%
{\sigma-k^{+}}\right]  .
\end{equation}
The corresponding subtraction term comes from the following terms in the LO
kernel (\ref{evolutionEq1})%
\begin{equation}
\frac{\alpha_{s}}{\pi^{2}}\int d\vec{z}_{4}U_{4}^{ba}\left\{  \frac{(\vec
{z}_{41}\vec{z}_{42})}{\vec{z}_{41}^{\,\,2}\vec{z}_{42}^{\,\,2}}(U_{1}%
t^{a})\cdot(t^{b}U_{2})\cdot U_{3}+\frac{(\vec{z}_{43}\vec{z}_{42})}{\vec
{z}_{43}^{\,\,2}\vec{z}_{42}^{\,\,2}}U_{1}\cdot(t^{b}U_{2})\cdot(U_{3}%
t^{a})\right\}  .
\end{equation}
Rewriting $U_{4}^{ba}$ as the trace of the Wilson lines in the fundamental
representation (\ref{Uadjoint}) and dressing $U_{4},U_{4}^{\dag}$ and $U_{3}$
or $U_{1}$ we get in particular
\[
\langle K_{LO}\otimes B_{123}^{\eta}\rangle|_{8}=\frac{\alpha_{s}^{2}}{\pi
^{4}}\int_{\sigma_{1}}^{\sigma}\frac{dp^{+}}{p^{+}}\int d\vec{z}_{0}%
U_{0}^{c^{\prime}c}\int d\vec{z}_{4}2tr([t^{b}t^{c^{\prime}}-t^{c^{\prime}%
}t^{b}]U_{4}t^{a}U_{4}^{\dag})
\]%
\begin{equation}
\times\left\{  \frac{(\vec{z}_{04}\vec{z}_{03})}{\vec{z}_{04}^{\,\,2}\vec
{z}_{03}^{\,\,2}}\frac{(\vec{z}_{41}\vec{z}_{42})}{\vec{z}_{41}^{\,\,2}\vec
{z}_{42}^{\,\,2}}(U_{1}t^{a})\cdot(t^{b}U_{2})\cdot(U_{3}t^{c})+\frac{(\vec
{z}_{43}\vec{z}_{42})}{\vec{z}_{43}^{\,\,2}\vec{z}_{42}^{\,\,2}}\frac{(\vec
{z}_{04}\vec{z}_{01})}{\vec{z}_{04}^{\,\,2}\vec{z}_{01}^{\,\,2}}(U_{1}%
t^{c})\cdot(t^{b}U_{2})\cdot(U_{3}t^{a})\right\}  .
\end{equation}
One can trim it to be%
\[
\langle K_{LO}\otimes B_{123}^{\eta}\rangle|_{8}=\frac{\alpha_{s}^{2}}{\pi
^{4}}if^{bc^{\prime}a^{\prime}}(U_{1}t^{a})\cdot(t^{b}U_{2})\cdot(U_{3}%
t^{c})\int_{\sigma_{1}}^{\sigma}\frac{dp^{+}}{p^{+}}\int d\vec{z}_{4}%
U_{4}^{c^{\prime}c}\int d\vec{z}_{0}U_{0}^{a^{\prime}a}%
\]%
\begin{equation}
\times\left\{  \frac{(\vec{z}_{40}\vec{z}_{43})}{\vec{z}_{04}^{\,\,2}\vec
{z}_{43}^{\,\,2}}\frac{(\vec{z}_{01}\vec{z}_{02})}{\vec{z}_{01}^{\,\,2}\vec
{z}_{02}^{\,\,2}}-\frac{(\vec{z}_{43}\vec{z}_{42})}{\vec{z}_{43}^{\,\,2}%
\vec{z}_{42}^{\,\,2}}\frac{(\vec{z}_{04}\vec{z}_{01})}{\vec{z}_{04}%
^{\,\,2}\vec{z}_{01}^{\,\,2}}\right\}  .
\end{equation}
Then%
\[
\langle K_{NLO}\otimes B_{123}^{\eta}\rangle|_{8}=\frac{\partial}{\partial
\eta}\langle B_{123}^{\eta}\rangle|_{8}-\langle K_{LO}\otimes B_{123}^{\eta
}\rangle|_{8}%
\]%
\[
=-\frac{\alpha_{s}^{2}}{\pi^{4}}if^{a^{\prime}bc^{\prime}}(U_{1}t^{a}%
)\cdot(t^{b}U_{2})\cdot(U_{3}t^{c})\int U_{0}^{a^{\prime}a}\frac{d\vec{z}_{0}%
}{\vec{z}_{10}^{\,\,2}}\int U_{4}^{c^{\prime}c}\frac{d\vec{z}_{4}}{\vec
{z}_{34}^{\,\,2}}\int_{0}^{\sigma}\frac{dk^{+}}{\vec{z}_{04}^{\,\,2}}%
\]%
\[
\times\left[  \frac{\left(  \vec{z}_{10}\vec{z}_{34}\right)  \left[  \vec
{z}_{24}^{\,\,2}-\vec{z}_{02}^{\,\,2}\right]  }{2((\sigma-k^{+})\,\vec{z}%
_{42}^{\,\,2}+k^{+}\,\vec{z}_{02}^{\,\,2})}+\frac{(\vec{z}_{10}\vec{z}%
_{40})(\vec{z}_{24}\vec{z}_{34})}{k^{+}}\left\{  \frac{\sigma}{(\sigma
-k^{+})\,\vec{z}_{42}^{\,\,2}+k^{+}\,\vec{z}_{02}^{\,\,2}}-\frac{1}{\vec
{z}_{42}^{\,\,2}}\right\}  \right.
\]%
\begin{equation}
\left.  -\frac{(\vec{z}_{04}\vec{z}_{34})(\vec{z}_{10}\vec{z}_{20})}%
{(\sigma-k^{+})}\left\{  \frac{\sigma}{(\sigma-k^{+})\,\vec{z}_{42}%
^{\,\,2}+k^{+}\,\vec{z}_{02}^{\,\,2}}-\frac{1}{\vec{z}_{02}^{\,\,2}}\right\}
\right]  .
\end{equation}
Thus we get the NLO contribution of diagram 8%
\[
\langle K_{NLO}\otimes B_{123}^{\eta}\rangle|_{8}=\frac{\alpha_{s}^{2}}%
{\pi^{4}}if^{a^{\prime}bc^{\prime}}(U_{1}t^{a})\cdot(t^{b}U_{2})\cdot
(U_{3}t^{c})\int U_{0}^{a^{\prime}a}d\vec{z}_{0}\int U_{4}^{c^{\prime}c}%
d\vec{z}_{4}%
\]%
\begin{equation}
\times\left[  \frac{1}{2\vec{z}_{04}^{\,\,2}}\frac{\left(  \vec{z}_{10}\vec
{z}_{34}\right)  }{\vec{z}_{10}^{\,\,2}\vec{z}_{34}^{\,\,2}}+\frac{(\vec
{z}_{10}\vec{z}_{40})}{\vec{z}_{10}^{\,\,2}\vec{z}_{40}^{\,\,2}}\frac{(\vec
{z}_{24}\vec{z}_{34})}{\vec{z}_{24}^{\,\,2}\vec{z}_{34}^{\,\,2}}+\frac
{(\vec{z}_{04}\vec{z}_{34})}{\vec{z}_{04}^{\,\,2}\vec{z}_{34}^{\,\,2}}%
\frac{(\vec{z}_{10}\vec{z}_{20})}{\vec{z}_{10}^{\,\,2}\vec{z}_{20}^{\,\,2}%
}\right]  \ln\frac{\vec{z}_{02}^{\,\,2}}{\vec{z}_{24}^{\,\,2}}.
\end{equation}
The contribution of the diagram which is a mirror reflection of diagram 8 with
respect to the shockwave reads%
\[
\langle B_{123}^{\eta}\rangle|_{8m}=-g^{4}f^{ab^{\prime}c}(t^{a^{\prime}}%
U_{1})\cdot(U_{2}t^{b})\cdot(t^{c^{\prime}}U_{3})\int_{0}^{\infty}dz_{1}%
^{+}\int_{0}^{\infty}dz_{3}^{+}\int_{-\infty}^{0}dz_{2}^{+}\int\theta
(-x^{+})d^{4}x
\]%
\[
\times\left\{  \frac{\partial G^{a^{\prime}a}(z_{1},x)_{\,\,\,\,\,j}^{-}%
}{\partial x^{\mu}}\left[  G_{0}^{b^{\prime}b}(x,z_{2})^{\mu-}G^{c^{\prime}%
c}(z_{3},x)^{-j}-G_{0}^{b^{\prime}b}(x,z_{2})^{j-}G^{c^{\prime}c}%
(z_{3},x)^{-\mu}\right]  \right.
\]%
\[
+\frac{\partial G_{0}^{b^{\prime}b}(x,z_{2})_{j}^{\,\,\,\,\,-}}{\partial
x^{\mu}}\left[  G^{a^{\prime}a}(z_{1},x)^{-j}G^{c^{\prime}c}(z_{3},x)^{-\mu
}-G^{a^{\prime}a}(z_{1},x)^{-\mu}G^{c^{\prime}c}(z_{3},x)^{-j}\right]
\]%
\begin{equation}
\left.  +\frac{\partial G^{c^{\prime}c}(z_{3},x)_{\,\,\,\,\,j}^{-}}{\partial
x^{\mu}}\left[  G^{a^{\prime}a}(z_{1},x)^{-\mu}G_{0}^{b^{\prime}b}%
(x,z_{2})^{j-}-G^{a^{\prime}a}(z_{1},x)^{-j}G_{0}^{b^{\prime}b}(x,z_{2}%
)^{\mu-}\right]  \right\}  .
\end{equation}
\[
\langle B_{123}^{\eta}\rangle|_{8m}=g^{4}f^{abc}(t^{a^{\prime}}U_{1}%
)\cdot(U_{2}t^{b})\cdot(t^{c^{\prime}}U_{3})\int_{-\infty}^{0}\frac{dz_{2}%
^{+}}{2(z_{2x}^{+})^{2}}\int\frac{d^{4}x}{(x^{+})^{3}}\theta(-z_{2x}^{+})
\]%
\[
\times\int_{\sigma_{1}}^{\sigma}\frac{dp^{+}}{(2\pi)^{2}} \int_{\sigma_{1}%
}^{\sigma}\frac{dk^{+}}{\left(  2\pi\right)  ^{3}}\int_{\sigma_{1}}^{\sigma
}\frac{dq^{+}}{\left(  2\pi\right)  ^{3}}e^{-iq^{+}\frac{\vec{z}_{x4}%
^{\,\,2}+i0}{2x^{+}}}e^{-ik^{+}\frac{\vec{z}_{x0}^{\,\,2}+i0}{2x^{+}}%
}e^{-ip^{+}\frac{\vec{z}_{2x}^{\,\,2}+i0}{2z_{2x}^{+}}}
\]
\[
\times\int U_{0}^{a^{\prime}a}\frac{d\vec{z}_{0}}{\vec{z}_{10}^{\,\,2}}\int
U_{4}^{c^{\prime}c}\frac{d\vec{z}_{4}}{\vec{z}_{34}^{\,\,2}}e^{-ix^{-}\left[
p^{+}-q^{+}-k^{+}\right]  }%
\]%
\[
\times\left[  k^{+}(\vec{z}_{10}\vec{z}_{2x})(\vec{z}_{34}\vec{z}_{4x}%
)+p^{+}\left[  (\vec{z}_{10}\vec{z}_{2x})(\vec{z}_{34}\vec{z}_{4x})-(\vec
{z}_{2x}\vec{z}_{34})\left(  \vec{z}_{10}\vec{z}_{0x}\right)  \right]
-q^{+}\left(  \vec{z}_{10}\vec{z}_{0x}\right)  (\vec{z}_{2x}\vec{z}_{34}%
)\frac{{}}{{}}\right.
\]%
\begin{equation}
\left.  +\frac{{}}{{}}k^{+}\left[  -\left(  \vec{z}_{10}\vec{z}_{34}\right)
(\vec{z}_{2x}\vec{z}_{x0})+\left(  \vec{z}_{10}\vec{z}_{2x}\right)  (\vec
{z}_{x0}\vec{z}_{34})\right]  +q^{+}\left[  -(\vec{z}_{2x}\vec{z}_{34}%
)(\vec{z}_{10}\vec{z}_{x4})+(\vec{z}_{2x}\vec{z}_{x4})\left(  \vec{z}_{10}%
\vec{z}_{34}\right)  \right]  \right]  .
\end{equation}
One can see that the structure in the brackets does not change and the
integration with respect to $z_{2x}^{+}$ gives the same contribution as for
diagram 8 while the integration with respect to $x^{+}$ gives the same
contribution with the opposite sign. Therefore the result for the diagram
which is a mirror reflection of diagram 8 with respect to the shockwave reads%
\[
\langle K_{NLO}\otimes B_{123}^{\eta}\rangle|_{8m}=-\frac{\alpha_{s}^{2}}%
{\pi^{4}}if^{abc}(t^{a^{\prime}}U_{1})\cdot(U_{2}t^{b})\cdot(t^{c^{\prime}%
}U_{3})\int U_{0}^{a^{\prime}a}d\vec{z}_{0}\int U_{4}^{c^{\prime}c}d\vec
{z}_{4}%
\]%
\begin{equation}
\times\left[  \frac{1}{2\vec{z}_{04}^{\,\,2}}\frac{\left(  \vec{z}_{10}\vec
{z}_{34}\right)  }{\vec{z}_{10}^{\,\,2}\vec{z}_{34}^{\,\,2}}+\frac{(\vec
{z}_{10}\vec{z}_{40})}{\vec{z}_{10}^{\,\,2}\vec{z}_{40}^{\,\,2}}\frac{(\vec
{z}_{24}\vec{z}_{34})}{\vec{z}_{24}^{\,\,2}\vec{z}_{34}^{\,\,2}}+\frac
{(\vec{z}_{04}\vec{z}_{34})}{\vec{z}_{04}^{\,\,2}\vec{z}_{34}^{\,\,2}}%
\frac{(\vec{z}_{10}\vec{z}_{20})}{\vec{z}_{10}^{\,\,2}\vec{z}_{20}^{\,\,2}%
}\right]  \ln\frac{\vec{z}_{02}^{\,\,2}}{\vec{z}_{24}^{\,\,2}}.
\end{equation}
Adding the contribution of diagram 8 we get%
\[
\langle K_{NLO}\otimes B_{123}^{\eta}\rangle|_{8+8m}=\frac{\alpha_{s}^{2}}%
{\pi^{4}}\int U_{0}^{a^{\prime}a}d\vec{z}_{0}\int U_{4}^{c^{\prime}c}d\vec
{z}_{4}%
\]%
\[
\times\left[  \frac{1}{2\vec{z}_{04}^{\,\,2}}\frac{\left(  \vec{z}_{10}\vec
{z}_{34}\right)  }{\vec{z}_{10}^{\,\,2}\vec{z}_{34}^{\,\,2}}+\frac{(\vec
{z}_{10}\vec{z}_{40})}{\vec{z}_{10}^{\,\,2}\vec{z}_{40}^{\,\,2}}\frac{(\vec
{z}_{24}\vec{z}_{34})}{\vec{z}_{24}^{\,\,2}\vec{z}_{34}^{\,\,2}}+\frac
{(\vec{z}_{04}\vec{z}_{34})}{\vec{z}_{04}^{\,\,2}\vec{z}_{34}^{\,\,2}}%
\frac{(\vec{z}_{10}\vec{z}_{20})}{\vec{z}_{10}^{\,\,2}\vec{z}_{20}^{\,\,2}%
}\right]  \ln\frac{\vec{z}_{02}^{\,\,2}}{\vec{z}_{24}^{\,\,2}}%
\]%
\begin{equation}
\times i\left\{  f^{a^{\prime}bc^{\prime}}(U_{1}t^{a})\cdot(t^{b}U_{2}%
)\cdot(U_{3}t^{c})-f^{abc}(t^{a^{\prime}}U_{1})\cdot(U_{2}t^{b})\cdot
(t^{c^{\prime}}U_{3})\right\}  .
\end{equation}
Combining this result with the contribution of diagram 7 and all the diagrams
obtained from it (\ref{diagram7+all}), we get the complete contribution of all
diagrams with two gluon intersecting the shockwave to the connected part of
the NLO kernel%
\[
\langle K_{NLO}^{conn}\otimes B_{123}^{\eta}\rangle|_{2g}=\frac{\alpha_{s}%
^{2}}{\pi^{4}}i\left\{  f^{a^{\prime}bc^{\prime}}(U_{1}t^{a})\cdot(t^{b}%
U_{2})\cdot(U_{3}t^{c})-f^{abc}(t^{a^{\prime}}U_{1})\cdot(U_{2}t^{b}%
)\cdot(t^{c^{\prime}}U_{3})\right\}
\]%
\[
\times\int d\vec{z}_{0}U_{0}^{a^{\prime}a}\int d\vec{z}_{4}U_{4}^{c^{\prime}c}%
\]%
\[
\times\left[  \frac{1}{2\vec{z}_{04}^{\,\,2}}\frac{\left(  \vec{z}_{10}\vec
{z}_{34}\right)  }{\vec{z}_{10}^{\,\,2}\vec{z}_{34}^{\,\,2}}+\frac{(\vec
{z}_{10}\vec{z}_{40})}{\vec{z}_{10}^{\,\,2}\vec{z}_{40}^{\,\,2}}\frac{(\vec
{z}_{24}\vec{z}_{34})}{\vec{z}_{24}^{\,\,2}\vec{z}_{34}^{\,\,2}}+\frac
{(\vec{z}_{04}\vec{z}_{34})}{\vec{z}_{04}^{\,\,2}\vec{z}_{34}^{\,\,2}}%
\frac{(\vec{z}_{10}\vec{z}_{20})}{\vec{z}_{10}^{\,\,2}\vec{z}_{20}^{\,\,2}%
}-\frac{\left(  \vec{z}_{20}\vec{z}_{10}\right)  }{\vec{z}_{02}{}^{2}\vec
{z}_{01}{}^{2}}\frac{\left(  \vec{z}_{24}\vec{z}_{34}\right)  }{\vec{z}%
_{24}^{\,\,2}\vec{z}_{34}{}^{2}}\right]  \ln\frac{\vec{z}_{02}^{\,\,2}}%
{\vec{z}_{24}^{\,\,2}}%
\]%
\begin{equation}
+(2\leftrightarrow1)+(2\leftrightarrow3).
\end{equation}
Using (\ref{Uadjoint}), one can rewrite it as
\[
\langle K_{NLO}^{conn}\otimes B_{123}^{\eta}\rangle|_{2g}=\frac{\alpha_{s}%
^{2}}{4\pi^{4}}\int d\vec{z}_{0}\int d\vec{z}_{4}\left\{  (U_{2}U_{0}^{\dag
}U_{1})\cdot U_{4}\cdot(U_{0}U_{4}^{\dag}U_{3})\right.
\]%
\[
\left.  +(U_{3}U_{4}^{\dag}U_{0})\cdot U_{4}\cdot(U_{1}U_{0}^{\dag}%
U_{2})-\frac{{}}{{}}\left(  1\leftrightarrow3,0\leftrightarrow4\right)
\right\}
\]%
\[
\times\left[  \frac{1}{2\vec{z}_{04}^{\,\,2}}\frac{\left(  \vec{z}_{10}\vec
{z}_{34}\right)  }{\vec{z}_{10}^{\,\,2}\vec{z}_{34}^{\,\,2}}+\frac{(\vec
{z}_{10}\vec{z}_{40})}{\vec{z}_{10}^{\,\,2}\vec{z}_{40}^{\,\,2}}\frac{(\vec
{z}_{24}\vec{z}_{34})}{\vec{z}_{24}^{\,\,2}\vec{z}_{34}^{\,\,2}}+\frac
{(\vec{z}_{04}\vec{z}_{34})}{\vec{z}_{04}^{\,\,2}\vec{z}_{34}^{\,\,2}}%
\frac{(\vec{z}_{10}\vec{z}_{20})}{\vec{z}_{10}^{\,\,2}\vec{z}_{20}^{\,\,2}%
}-\frac{\left(  \vec{z}_{20}\vec{z}_{10}\right)  }{\vec{z}_{02}{}^{2}\vec
{z}_{01}{}^{2}}\frac{\left(  \vec{z}_{24}\vec{z}_{34}\right)  }{\vec{z}%
_{24}^{\,\,2}\vec{z}_{34}{}^{2}}\right]  \ln\frac{\vec{z}_{02}^{\,\,2}}%
{\vec{z}_{24}^{\,\,2}}%
\]%
\begin{equation}
+(2\leftrightarrow1)+(2\leftrightarrow3). \label{2gResultCoordinate}%
\end{equation}

\section{Diagrams with 1 gluon intersecting the shockwave}%
\begin{figure}
[ptbh]
\begin{center}
\includegraphics[
width=\textwidth
]%
{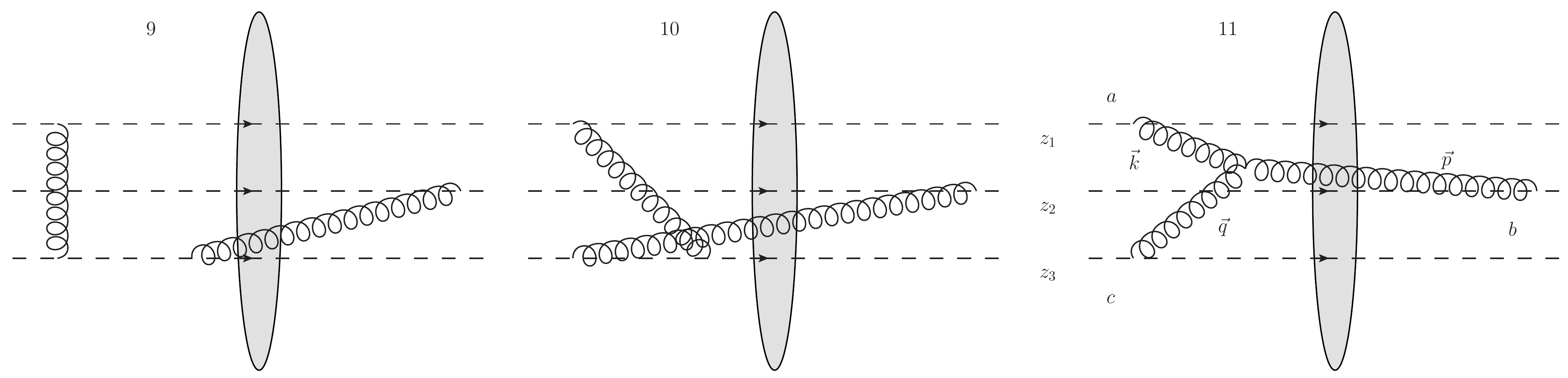}%
\caption{Diagrams with one gluon intersecting the shockwave.}%
\label{1gluoninSW}%
\end{center}
\end{figure}
Let us turn to the diagrams in fig. \ref{1gluoninSW}. Hereafter we will not
write the LO subtraction terms explicitly. Instead we will set $\sigma_{1}=0$
and use the $\left[  \frac{1}{p^{+}}\right]  _{+}$ and $\left[  \frac
{1}{\sigma-p^{+}}\right]  _{+}$ prescriptions where necessary%
\begin{equation}
\int_{0}^{\sigma}dp^{+}f(p^{+})\left[  \frac{1}{p^{+}}\right]  _{+}=\int
_{0}^{\sigma}dp^{+}\frac{f(p^{+})-f(0)}{p^{+}},
\end{equation}%
\begin{equation}
\int_{0}^{\sigma}dp^{+}f(p^{+})\left[  \frac{1}{\sigma-p^{+}}\right]
_{+}=\int_{0}^{\sigma}dp^{+}\frac{f(p^{+})-f(\sigma)}{\sigma-p^{+}}.
\end{equation}
Diagram 9 reads%
\[
\langle B_{123}^{\eta}\rangle|_{9}=g^{4}(U_{1}t^{a})\cdot(t^{b^{\prime}}%
U_{2})\cdot(U_{3}t^{b}t^{a^{\prime}})
\]%
\begin{equation}
\times\int_{-\infty}^{0}dz_{1}^{+}\int_{0}^{\infty}dz_{2}^{+}\int_{-\infty
}^{0}dz_{3}^{+}\int_{z_{3}^{+}}^{0}dz_{3}^{\prime+}G^{--}(z_{2},z_{3}^{\prime
})^{b^{\prime}b}G^{--}(z_{3},z_{1})^{a^{\prime}a}.
\end{equation}
Using the momentum representation of the propagators (\ref{G--}) and
(\ref{GinSWmom}) and as in the LO\ calculation changing $-ik^{-}z_{31}%
^{+}\rightarrow-i(k^{-}+i\varepsilon)z_{3}^{+}+i(k^{-}-i\varepsilon)z_{1}^{+}$
and integrating first with respect to $z_{1}^{+},$ $z_{3}^{+}$ and then to
$k^{-}$ next, we get%
\[
\langle B_{123}^{\eta}\rangle|_{9}=g^{4}(U_{1}t^{a})\cdot(t^{b^{\prime}}%
U_{2})\cdot(U_{3}t^{b}t^{a})\int_{0}^{\sigma}\frac{dp^{+}}{p^{+}\left(
2\pi\right)  ^{2}}\int\frac{(z_{20})_{\bot}^{\alpha}}{\vec{z}_{20}^{\,\,\,2}%
}U_{0}^{b^{\prime}b}d\vec{z}_{0}\int_{-\infty}^{0}dz_{3}^{\prime+}%
\]%
\begin{equation}
\times\int\frac{d\vec{p}}{\left(  2\pi\right)  ^{2}}e^{i\vec{p}\vec{z}_{03}%
}e^{i\frac{z_{3}^{\prime+}}{2p^{+}}\left(  \vec{p}^{\,\,2}-i0\right)  }%
\frac{p_{\bot\alpha}}{p^{+}}\int\frac{dk^{+}}{2\pi k^{+}}\int\frac{d\vec{k}%
}{(2\pi)^{2}}e^{i\,\vec{k}\,\vec{z}_{31}}\frac{2\theta\left(  -k^{+}\right)
}{\vec{k}^{\,\,2}-i0}e^{-i\frac{\vec{k}^{\,\,2}-i0}{2k^{+}}z_{3}^{\prime+}}.
\end{equation}
Then one can integrate w.r.t. $z_{3}^{\prime+}$%
\[
\langle B_{123}^{\eta}\rangle|_{9}=2ig^{4}(U_{1}t^{a})\cdot(t^{b^{\prime}%
}U_{2})\cdot(U_{3}t^{b}t^{a})\int_{0}^{\sigma}\frac{dp^{+}}{p^{+}\left(
2\pi\right)  ^{7}}\int\frac{(z_{20})_{\bot}^{\alpha}}{\vec{z}_{20}^{\,\,\,2}%
}U_{0}^{b^{\prime}b}d\vec{z}_{0}%
\]%
\begin{equation}
\times\int d\vec{p}e^{i\vec{p}\vec{z}_{03}+i\,\vec{k}\,\vec{z}_{31}}%
p_{\bot\alpha}\int\frac{d\vec{k}}{\vec{k}^{\,\,2}}\int_{0}^{\sigma}dk^{+}%
\frac{2}{\vec{p}^{\,\,2}k^{+}+\vec{k}^{\,\,2}p^{+}}.
\end{equation}
Therefore using the $\left[  \frac{1}{p^{+}}\right]  _{+}$ prescription one
gets%
\begin{equation}
\langle K_{NLO}\otimes B_{123}^{\eta}\rangle|_{9}=4ig^{4}(U_{1}t^{a}%
)\cdot(t^{b^{\prime}}U_{2})\cdot(U_{3}t^{b}t^{a})
\end{equation}%
\[
\times\int\frac{(z_{20})_{\bot}^{\alpha}}{\left(  2\pi\right)  ^{7}\vec
{z}_{20}^{\,\,\,2}}U_{0}^{b^{\prime}b}d\vec{z}_{0}\int d\vec{p}e^{i\vec{p}%
\vec{z}_{03}+i\,\vec{k}\,\vec{z}_{31}}\int\frac{d\vec{k}}{\vec{k}^{\,\,2}%
}\frac{p_{\bot\alpha}}{\vec{p}^{\,\,2}}\ln\frac{\vec{p}^{\,\,2}}{\vec
{k}^{\,\,2}}.
\]
Now we consider diagram 10. It has the form%
\[
\langle B_{123}^{\eta}\rangle|_{10}=g^{4}(U_{1}t^{a})\cdot(t^{b^{\prime}}%
U_{2})\cdot(U_{3}t^{a^{\prime}}t^{b})
\]%
\begin{equation}
\times\int_{-\infty}^{0}dz_{1}^{+}\int_{0}^{\infty}dz_{2}^{+}\int_{-\infty
}^{0}dz_{3}^{\prime+}\int_{z_{3}^{\prime+}}^{0}dz_{3}^{+}G^{--}(z_{2}%
,z_{3}^{\prime})^{b^{\prime}b}G^{--}(z_{3},z_{1})^{a^{\prime}a}.
\end{equation}
Going through the same steps as for diagram 9 one gets%
\[
\langle K_{NLO}\otimes B_{123}^{\eta}\rangle|_{10}=4ig^{4}(U_{1}t^{a}%
)\cdot(t^{b^{\prime}}U_{2})\cdot(U_{3}t^{a}t^{b})
\]%
\begin{equation}
\times\int\frac{(z_{20})_{\bot}^{\alpha}}{\left(  2\pi\right)  ^{7}\vec
{z}_{20}^{\,\,\,2}}U_{0}^{b^{\prime}b}d\vec{z}_{0}\int\frac{d\vec{k}}{\vec
{k}^{\,\,2}}\int d\vec{p}e^{i\,\vec{k}\,\vec{z}_{31}+i\vec{p}\vec{z}_{03}%
}\frac{p_{\bot\alpha}}{\vec{p}^{\,\,2}}\ln\frac{\vec{k}^{2}}{\vec{p}^{\,\,2}}.
\end{equation}
Then the contribution of both diagrams and the diagrams with the
$3\leftrightarrow1$ Wilson line substitution reads%
\[
\langle K_{NLO}\otimes B_{123}^{\eta}\rangle|_{9+10+(9+10)(3\leftrightarrow
1)}=4f^{abc}g^{4}(U_{1}t^{a})\cdot(t^{b^{\prime}}U_{2})\cdot(U_{3}t^{c}%
)\int\frac{(z_{20})_{\bot}^{\alpha}}{\left(  2\pi\right)  ^{7}\vec{z}%
_{20}^{\,\,\,2}}U_{0}^{b^{\prime}b}d\vec{z}_{0}%
\]%
\begin{equation}
\times\int d\vec{p}\int\frac{d\vec{k}}{\vec{k}^{\,\,2}}\frac{p_{\bot\alpha}%
}{\vec{p}^{\,\,2}}\ln\frac{\vec{p}^{\,\,2}}{\vec{k}^{\,\,2}}\left(
e^{i\vec{p}\vec{z}_{03}+i\,\vec{k}\,\vec{z}_{31}}-e^{i\vec{p}\vec{z}%
_{01}-i\,\vec{k}\,\vec{z}_{31}}\right)  . \label{9+10+(3<->1)}%
\end{equation}
We turn to diagram 11.
\[
\langle B_{123}^{\eta}\rangle|_{11}=-g^{4}f^{a^{\prime}b^{\prime}c^{\prime}%
}(U_{1}t^{a})\cdot(t^{b}U_{2})\cdot(U_{3}t^{c})\int_{-\infty}^{0}dz_{1}%
^{+}\int_{-\infty}^{0}dz_{3}^{+}\int_{0}^{\infty}dz_{2}^{+}\int\theta
(-x^{+})d^{4}x
\]%
\[
\times\left\{  \frac{\partial G_{0}^{a^{\prime}a}(x,z_{1})_{j}^{\,\,\,\,\,-}%
}{\partial x^{\mu}}\left[  G^{bb^{\prime}}(z_{2},x)^{-\mu}G_{0}^{c^{\prime}%
c}(x,z_{3})^{j-}-G^{bb^{\prime}}(z_{2},x)^{-j}G_{0}^{c^{\prime}c}%
(x,z_{3})^{\mu-}\right]  \right.
\]%
\[
+\frac{\partial G^{bb^{\prime}}(z_{2},x)_{\,\,\,j}^{-}}{\partial x^{\mu}%
}\left[  G_{0}^{a^{\prime}a}(x,z_{1})^{j-}G_{0}^{c^{\prime}c}(x,z_{3})^{\mu
-}-G_{0}^{a^{\prime}a}(x,z_{1})^{\mu-}G_{0}^{c^{\prime}c}(x,z_{3}%
)^{j-}\right]
\]%
\begin{equation}
\left.  +\frac{\partial G_{0}^{c^{\prime}c}(x,z_{3})_{j}^{\,\,\,\,\,-}%
}{\partial x^{\mu}}\left[  G_{0}^{a^{\prime}a}(x,z_{1})^{\mu-}G^{bb^{\prime}%
}(z_{2},x)^{-j}-G_{0}^{a^{\prime}a}(x,z_{1})^{j-}G^{bb^{\prime}}%
(z_{2},x)^{-\mu}\right]  \right\}  .
\end{equation}
Here we sum over $j=1,2$ and $\mu=-,1,2.$ It is convenient to split this
expression into two parts.%
\begin{equation}
\langle B_{123}^{\eta}\rangle|_{11}=\langle B_{123}^{\eta}\rangle|_{11_{1}%
}+\langle B_{123}^{\eta}\rangle|_{11_{2}}.
\end{equation}%
\[
\langle B_{123}^{\eta}\rangle|_{11_{1}}=-g^{4}f^{a^{\prime}b^{\prime}%
c^{\prime}}(U_{1}t^{a})\cdot(t^{b}U_{2})\cdot(U_{3}t^{c})\int_{-\infty}%
^{0}dz_{1}^{+}\int_{-\infty}^{0}dz_{3}^{+}\int_{0}^{\infty}dz_{2}^{+}%
\int\theta(-x^{+})d^{4}x
\]%
\[
\times\left\{  G_{0}^{c^{\prime}c}(x,z_{3})^{--}\left[  G_{0}^{a^{\prime}%
a}(x,z_{1})^{j-}\frac{\partial G^{bb^{\prime}}(z_{2},x)_{\,\,\,j}^{-}%
}{\partial x^{-}}-G^{bb^{\prime}}(z_{2},x)^{-j}\frac{\partial G_{0}%
^{a^{\prime}a}(x,z_{1})_{j}^{\,\,\,\,\,-}}{\partial x^{-}}\right]  \right.
\]%
\begin{equation}
\left.  +G_{0}^{a^{\prime}a}(x,z_{1})^{--}\left[  G^{bb^{\prime}}%
(z_{2},x)^{-j}\frac{\partial G_{0}^{c^{\prime}c}(x,z_{3})_{j}^{\,\,\,\,\,-}%
}{\partial x^{-}}-G_{0}^{c^{\prime}c}(x,z_{3})^{j-}\frac{\partial
G^{bb^{\prime}}(z_{2},x)_{\,\,\,j}^{-}}{\partial x^{-}}\right]  \right\}  .
\end{equation}%
\[
\langle B_{123}^{\eta}\rangle|_{11_{1}}=-g^{4}f^{a^{\prime}b^{\prime}%
c^{\prime}}(U_{1}t^{a})\cdot(t^{b}U_{2})\cdot(U_{3}t^{c})\int_{-\infty}%
^{0}dz_{1}^{+}\int_{-\infty}^{0}dz_{3}^{+}\int_{0}^{\infty}dz_{2}^{+}%
\int\theta(-x^{+})d^{4}x
\]%
\[
\times\left\{  \int\frac{dq^{+}}{2\pi}e^{-ix^{-}q^{+}}\int\frac{d\vec{q}%
dq^{-}}{(2\pi)^{3}}\frac{2iq^{-}e^{-iq^{-}x^{+}+iq^{-}z_{3}^{+}+i\,\vec
{q}\,\vec{z}_{x3}}}{q^{+}(q^{2}+i0)}\right.
\]%
\[
\times\left[  G_{0}^{a^{\prime}a}(x,z_{1})^{j-}\frac{\partial G^{bb^{\prime}%
}(z_{2},x)_{\,\,\,j}^{-}}{\partial x^{-}}-G^{bb^{\prime}}(z_{2},x)^{-j}%
\frac{\partial G_{0}^{a^{\prime}a}(x,z_{1})_{j}^{\,\,\,\,\,-}}{\partial x^{-}%
}\right]  +\int\frac{dk^{+}}{2\pi}e^{-ix^{-}k^{+}}\int\frac{d\vec{k}dk^{-}%
}{(2\pi)^{3}}%
\]%
\begin{equation}
\times\left.  \frac{2ik^{-}e^{-ik^{-}x^{+}+ik^{-}z_{1}^{+}+i\,\vec{k}\,\vec
{z}_{x1}}}{k^{+}(k^{2}+i0)}\left[  G^{bb^{\prime}}(z_{2},x)^{-j}\frac{\partial
G_{0}^{c^{\prime}c}(x,z_{3})_{j}^{\,\,\,\,\,-}}{\partial x^{-}}-G_{0}%
^{c^{\prime}c}(x,z_{3})^{j-}\frac{\partial G^{bb^{\prime}}(z_{2}%
,x)_{\,\,\,j}^{-}}{\partial x^{-}}\right]  \right\}  .
\end{equation}
Substituting the propagators and using $\left[  \frac{1}{\sigma-p^{+}}\right]
_{+}$ prescription one gets%
\[
\frac{\partial}{\partial\eta}\langle B_{123}^{\eta}\rangle|_{11_{1}}%
=-2g^{4}f^{a^{\prime}b^{\prime}c^{\prime}}(U_{1}t^{a})\cdot(t^{b}U_{2}%
)\cdot(U_{3}t^{c})\int\frac{d\vec{z}_{0}}{\left(  2\pi\right)  ^{7}}%
U_{0}^{bb^{\prime}}\frac{\left(  z_{20}\right)  _{\bot}^{\alpha}}{\vec{z}%
_{20}^{\,\,2}}\int d\vec{k}\int d\vec{q}\int_{0}^{\sigma}dp^{+}%
\]%
\[
\times e^{-i\,\vec{k}\,\vec{z}_{10}-i\,\vec{q}\,\vec{z}_{30}}\left[  \frac
{-1}{(\sigma-p^{+})(\vec{q}+\vec{k})^{2}+p^{+}\vec{q}^{\,\,2}}\frac
{k_{\alpha\bot}}{\vec{k}^{2}}+\frac{1}{\left(  \sigma-p^{+}\right)  (\vec
{q}+\vec{k})^{2}+p^{+}\vec{k}^{\,\,2}}\frac{q_{\alpha\bot}}{\vec{q}^{\,\,2}%
}\right.
\]%
\begin{equation}
\left.  +\frac{2(\vec{q}^{\,\,2}-(\vec{q}+\vec{k})^{2})}{\left[  (\sigma
-p^{+})(\vec{q}+\vec{k})^{2}+p^{+}\vec{q}^{\,\,2}\right]  \vec{q}^{\,\,2}%
}\frac{k_{\alpha\bot}}{\vec{k}^{2}}-\frac{2(\vec{k}^{\,\,2}-(\vec{q}+\vec
{k})^{2})}{\left[  \left(  \sigma-p^{+}\right)  (\vec{q}+\vec{k})^{2}%
+p^{+}\vec{k}^{\,\,2}\right]  \vec{k}^{\,\,2}}\frac{q_{\alpha\bot}}{\vec
{q}^{\,\,2}}\right]  .
\end{equation}
Integrating with respect to $p^{+}$ one comes to
\[
\langle K_{NLO}\otimes B_{123}^{\eta}\rangle|_{11_{1}}=-2g^{4}f^{ab^{\prime}%
c}(U_{1}t^{a})\cdot(t^{b}U_{2})\cdot(U_{3}t^{c})\int\frac{d\vec{z}_{0}%
}{\left(  2\pi\right)  ^{7}}U_{0}^{bb^{\prime}}\frac{\left(  z_{20}\right)
_{\bot}^{\alpha}}{\vec{z}_{20}^{\,\,2}}\int d\vec{k}\int d\vec{q}%
\]%
\[
\times e^{-i\,\vec{k}\,\vec{z}_{10}-i\,\vec{q}\,\vec{z}_{30}}\left[
\frac{q_{\alpha\bot}}{\vec{q}^{\,\,2}}\frac{1}{\vec{k}^{\,\,2}-(\vec{q}%
+\vec{k})^{2}}\ln\frac{\vec{k}^{\,\,2}}{(\vec{q}+\vec{k})^{2}}-\frac
{k_{\alpha\bot}}{\vec{k}^{2}}\frac{1}{\vec{q}^{\,\,2}-(\vec{q}+\vec{k})^{2}%
}\ln\frac{\vec{q}^{\,\,2}}{(\vec{q}+\vec{k})^{2}}\right.
\]%
\begin{equation}
\left.  +\frac{2}{\vec{q}^{\,\,2}}\frac{k_{\alpha\bot}}{\vec{k}^{2}}\ln
\frac{\vec{q}^{\,\,2}}{(\vec{q}+\vec{k})^{2}}-\frac{2}{\vec{k}^{\,\,2}}%
\ln\frac{\vec{k}^{\,\,2}}{(\vec{q}+\vec{k})^{2}}\frac{q_{\alpha\bot}}{\vec
{q}^{\,\,2}}\right]  .
\end{equation}
Adding this contribution to the the contribution of diagrams 9 and 10
(\ref{9+10+(3<->1)}), one gets the regular contribution%
\[
\langle K_{NLO}\otimes B_{123}^{\eta}\rangle|_{11_{1}%
+9+10+(9+10)(3\leftrightarrow1)}=-2f^{abc}g^{4}(U_{1}t^{a})\cdot(t^{b^{\prime
}}U_{2})\cdot(U_{3}t^{c})
\]%
\[
\times\int\frac{(z_{20})_{\bot}^{\alpha}}{\left(  2\pi\right)  ^{7}\vec
{z}_{20}^{\,\,\,2}}U_{0}^{b^{\prime}b}d\vec{z}_{0}\int d\vec{k}\int d\vec{q}%
\]%
\[
\times\left[  e^{-i\,\vec{k}\,\vec{z}_{10}-i\,\vec{q}\,\vec{z}_{30}}\left\{
-\frac{k_{\alpha\bot}}{\vec{k}^{2}}\frac{1}{\vec{q}^{\,\,2}-(\vec{q}+\vec
{k})^{2}}\ln\frac{\vec{q}^{\,\,2}}{(\vec{q}+\vec{k})^{2}}+\frac{q_{\alpha\bot
}}{\vec{q}^{2}}\frac{1}{\vec{k}^{\,\,2}-(\vec{q}+\vec{k})^{2}}\ln\frac{\vec
{k}^{\,\,2}}{(\vec{q}+\vec{k})^{2}}\right\}  \right.
\]%
\[
+2\frac{1}{\vec{q}^{\,\,2}}\frac{k_{\bot\alpha}}{\vec{k}^{2}}\left\{
e^{-i\,\vec{k}\,\vec{z}_{10}-i\,\vec{q}\,\vec{z}_{30}}\ln\frac{\vec{q}%
^{\,\,2}}{(\vec{q}+\vec{k})^{2}}+e^{-i\vec{k}\vec{z}_{10}-i\,\vec{q}\,\vec
{z}_{31}}\ln\frac{\vec{k}^{2}}{\vec{q}^{\,\,2}}\right\}
\]%
\begin{equation}
\left.  +2\frac{q_{\alpha\bot}}{\vec{q}^{2}}\frac{1}{\vec{k}^{\,\,2}}\left\{
-e^{-i\,\vec{k}\,\vec{z}_{10}-i\,\vec{q}\,\vec{z}_{30}}\ln\frac{\vec
{k}^{\,\,2}}{(\vec{q}+\vec{k})^{2}}-e^{-i\vec{q}\vec{z}_{30}+i\,\vec{k}%
\,\vec{z}_{31}}\ln\frac{\vec{q}^{\,\,2}}{\vec{k}^{\,\,2}}\right\}  \right]  .
\label{11-1+9+10+(3<->1)}%
\end{equation}
The second contribution to diagram 11 reads%
\[
\langle B_{123}^{\eta}\rangle|_{11_{2}}=-g^{4}f^{a^{\prime}b^{\prime}%
c^{\prime}}(U_{1}t^{a})\cdot(t^{b}U_{2})\cdot(U_{3}t^{c})\int_{-\infty}%
^{0}dz_{1}^{+}\int_{-\infty}^{0}dz_{3}^{+}\int_{0}^{\infty}dz_{2}^{+}%
\int\theta(-x^{+})d^{4}x
\]%
\[
\times\left\{  G^{bb^{\prime}}(z_{2},x)^{--}\left[  \frac{\partial
G_{0}^{a^{\prime}a}(x,z_{1})_{j}^{\,\,\,\,\,-}}{\partial x^{-}}G_{0}%
^{c^{\prime}c}(x,z_{3})^{j-}-G_{0}^{a^{\prime}a}(x,z_{1})^{j-}\frac{\partial
G_{0}^{c^{\prime}c}(x,z_{3})_{j}^{\,\,\,\,\,-}}{\partial x^{-}}\right]
\right.
\]%
\[
+\frac{\partial G_{0}^{a^{\prime}a}(x,z_{1})_{j}^{\,\,\,\,\,-}}{\partial
x^{l}}\left[  G^{bb^{\prime}}(z_{2},x)^{-l}G_{0}^{c^{\prime}c}(x,z_{3}%
)^{j-}-G^{bb^{\prime}}(z_{2},x)^{-j}G_{0}^{c^{\prime}c}(x,z_{3})^{l-}\right]
\]%
\[
+\frac{\partial G^{bb^{\prime}}(z_{2},x)_{\,\,\,j}^{-}}{\partial x^{l}}\left[
G_{0}^{a^{\prime}a}(x,z_{1})^{j-}G_{0}^{c^{\prime}c}(x,z_{3})^{l-}%
-G_{0}^{a^{\prime}a}(x,z_{1})^{l-}G_{0}^{c^{\prime}c}(x,z_{3})^{j-}\right]
\]%
\begin{equation}
\left.  +\frac{\partial G_{0}^{c^{\prime}c}(x,z_{3})_{j}^{\,\,\,\,\,-}%
}{\partial x^{l}}\left[  G_{0}^{a^{\prime}a}(x,z_{1})^{l-}G^{bb^{\prime}%
}(z_{2},x)^{-j}-G_{0}^{a^{\prime}a}(x,z_{1})^{j-}G^{bb^{\prime}}(z_{2}%
,x)^{-l}\right]  \right\}  .
\end{equation}
Substituting the propagators and integrating with respect to $z_{2}^{+},$and
$\vec{x},\vec{p}$ one obtains
\[
\langle B_{123}^{\eta}\rangle|_{11_{2}}=-g^{4}f^{a^{\prime}b^{\prime}%
c^{\prime}}(U_{1}t^{a})\cdot(t^{b}U_{2})\cdot(U_{3}t^{c})\int_{-\infty}%
^{0}dz_{1}^{+}\int_{-\infty}^{0}dz_{3}^{+}%
\]%
\[
\times\int\theta(-x^{+})dx^{+}dx^{-}\int d\vec{z}_{0}U_{0}^{bb^{\prime}}%
\frac{(z_{20})_{\bot}^{\alpha}}{\vec{z}_{20}^{\,\,2}}\int_{\sigma_{1}}%
^{\sigma}\frac{dp^{+}}{p^{+}\left(  2\pi\right)  ^{2}}e^{ip^{+}x^{-}}\int
\frac{dk^{+}}{2\pi}e^{-ik^{+}x^{-}}\int\frac{dq^{+}}{2\pi}e^{-iq^{+}x^{-}}%
\]%
\[
\times\int\frac{d\vec{k}}{\left(  2\pi\right)  ^{2}}e^{-i\,\vec{k}\,\vec
{z}_{1}}\frac{e^{i\frac{\vec{k}^{2}-i0}{2k^{+}}z_{1x}^{+}}}{2\left(
k^{+}\right)  ^{2}}\int\frac{d\vec{q}}{\left(  2\pi\right)  ^{2}}%
e^{-i\,\vec{q}\,\vec{z}_{3}}\frac{e^{i\frac{\vec{q}^{2}-i0}{2q^{+}}z_{3x}^{+}%
}}{2\left(  q^{+}\right)  ^{2}}e^{i(\vec{q}+\vec{k})\vec{z}_{0}}%
e^{i\frac{x^{+}}{2p^{+}}(\vec{q}+\vec{k})^{2}}%
\]%
\[
\times\left(  \theta(-z_{3x}^{+})\theta(q^{+})\theta(-z_{1x}^{+})\theta
(k^{+})-\theta(z_{3x}^{+})\theta(-q^{+})\theta(-z_{1x}^{+})\theta
(k^{+})-\theta(z_{1x}^{+})\theta(-k^{+})\theta(-z_{3x}^{+})\theta
(q^{+})\right)
\]%
\begin{equation}
\times\left\{  \frac{(-k-q)_{\alpha\bot}}{p^{+}}i\left[  k^{+}-q^{+}\right]
(\vec{k}\vec{q})-i\left[  k_{\alpha\bot}(\vec{q}^{\,\,2}+(\vec{k}\vec
{q}))-((\vec{q}\vec{k})+\vec{k}^{\,\,2})q_{\alpha\bot}\right]  \right\}  .
\end{equation}
Then one integrates with respect to $z_{1x}^{+}$ and $z_{3x}^{+}$ and changes
$k^{+}\rightarrow-k^{+}$ in the term with $\theta\left(  -k^{+}\right)  $ and
$q^{+}\rightarrow-q^{+}$ in the term with $\theta\left(  -q^{+}\right)  .$
After integrating with respect to $x^{+}$,$x^{-}$ and using the momentum
conservation law, one has for $\frac{\partial}{\partial\eta}\langle
B_{123}^{\eta}\rangle|_{11_{2}}$%
\[
\frac{\partial}{\partial\eta}\langle B_{123}^{\eta}\rangle|_{11_{2}}%
=2g^{4}f^{a^{\prime}b^{\prime}c^{\prime}}(U_{1}t^{a})\cdot(t^{b}U_{2}%
)\cdot(U_{3}t^{c})\int d\vec{z}_{0}U_{0}^{bb^{\prime}}\frac{(z_{20})_{\bot
}^{\alpha}}{\left(  2\pi\right)  ^{7}\vec{z}_{20}^{\,\,2}}\int\frac{d\vec{k}%
}{\vec{k}^{2}}\int\frac{d\vec{q}}{\vec{q}^{\,\,2}}e^{i(\vec{q}\vec{z}%
_{03}+\vec{k}\vec{z}_{01})}%
\]%
\[
\times\left(  \int_{0}^{\sigma}dp^{+}\frac{(k+q)_{\alpha\bot}(\vec{k}\vec
{q})-\left[  k_{\alpha\bot}(\vec{q}^{\,\,2}+(\vec{k}\vec{q}))-((\vec{q}\vec
{k})+\vec{k}^{\,\,2})q_{\alpha\bot}\right]  }{(\sigma-p^{+})(\vec{q}+\vec
{k})^{2}+p^{+}\vec{q}^{\,\,2}}\right.
\]%
\[
-2\sigma(\vec{k}\vec{q})(k+q)_{\alpha\bot}\int_{0}^{\sigma}\frac{dp^{+}}%
{p^{+}}\left[  \frac{1}{(\sigma-p^{+})(\vec{q}+\vec{k})^{2}+p^{+}\vec
{q}^{\,\,2}}-\frac{1}{\sigma(\vec{q}+\vec{k})^{2}}\right]
\]%
\[
+\int_{0}^{\sigma}dp^{+}\frac{-(k+q)_{\alpha\bot}(\vec{k}\vec{q})-\left[
k_{\alpha\bot}(\vec{q}^{\,\,2}+(\vec{k}\vec{q}))-((\vec{q}\vec{k})+\vec
{k}^{\,\,2})q_{\alpha\bot}\right]  }{\vec{k}^{2}p^{+}+(\sigma-p^{+})(\vec
{q}+\vec{k})^{2}}%
\]%
\begin{equation}
\left.  +2\sigma(k+q)_{\alpha\bot}(\vec{k}\vec{q})\int_{0}^{\sigma}%
\frac{dp^{+}}{p^{+}}\left[  \frac{1}{\vec{k}^{2}p^{+}+(\sigma-p^{+})(\vec
{q}+\vec{k})^{2}}-\frac{1}{\sigma(\vec{q}+\vec{k})^{2}}\right]  \right)  .
\end{equation}
As a result,%
\[
\langle K_{NLO}\otimes B_{123}^{\eta}\rangle|_{11_{2}}=2g^{4}f^{a^{\prime
}b^{\prime}c^{\prime}}(U_{1}t^{a})\cdot(t^{b}U_{2})\cdot(U_{3}t^{c})\int
d\vec{z}_{0}U_{0}^{bb^{\prime}}\frac{(z_{20})_{\bot}^{\alpha}}{\left(
2\pi\right)  ^{7}\vec{z}_{20}^{\,\,2}}\int d\vec{k}\int d\vec{q}e^{i(\vec
{q}\vec{z}_{03}+\vec{k}\vec{z}_{01})}%
\]%
\[
\times\left(  \frac{-k_{\alpha\bot}}{(\vec{q}^{\,\,2}-(\vec{q}+\vec{k}%
)^{2})\vec{k}^{2}}\ln\frac{\vec{q}^{\,\,2}}{(\vec{q}+\vec{k})^{2}}%
+\frac{q_{\alpha\bot}}{(\vec{k}^{2}-(\vec{q}+\vec{k})^{2})\vec{q}^{\,\,2}}%
\ln\frac{\vec{k}^{2}}{(\vec{q}+\vec{k})^{2}}\right.
\]%
\begin{equation}
-\frac{q_{\alpha\bot}}{\vec{q}^{\,\,2}\vec{k}^{\,\,2}}\ln\frac{\vec{q}%
^{\,\,2}}{(\vec{q}+\vec{k})^{2}}+\frac{k_{\alpha\bot}}{\vec{q}^{\,\,2}\vec
{k}^{\,\,2}}\ln\frac{\vec{k}^{2}}{(\vec{q}+\vec{k})^{2}}\left.  -\frac
{2(\vec{q}\vec{k})(k+q)_{\alpha\bot}}{\vec{q}^{\,\,2}\vec{k}^{\,\,2}(\vec
{q}+\vec{k})^{2}}\ln\frac{\vec{k}^{2}}{\vec{q}^{\,\,2}}\right)  .
\end{equation}
Adding to this expression the first contribution of diagram 11 and diagrams 9
and 10 with the corresponding (1$\leftrightarrow$3) symmetrization
(\ref{11-1+9+10+(3<->1)}) one has%
\[
\langle K_{NLO}\otimes B_{123}^{\eta}\rangle|_{11+9+10+(9+10)(3\leftrightarrow
1)}=2g^{4}f^{a^{\prime}b^{\prime}c^{\prime}}(U_{1}t^{a})\cdot(t^{b}U_{2}%
)\cdot(U_{3}t^{c})\int d\vec{z}_{0}U_{0}^{bb^{\prime}}\frac{(z_{20})_{\bot
}^{\alpha}}{\left(  2\pi\right)  ^{7}\vec{z}_{20}^{\,\,2}}%
\]%
\[
\times\int d\vec{k}\int d\vec{q}\left(  -2\frac{1}{\vec{q}^{\,\,2}}%
\frac{k_{\bot\alpha}}{\vec{k}^{2}}\left\{  e^{-i\,\vec{k}\,\vec{z}%
_{10}-i\,\vec{q}\,\vec{z}_{30}}\ln\frac{\vec{q}^{\,\,2}}{(\vec{q}+\vec{k}%
)^{2}}+e^{-i\vec{k}\vec{z}_{10}-i\,\vec{q}\,\vec{z}_{31}}\ln\frac{\vec{k}^{2}%
}{\vec{q}^{\,\,2}}\right\}  \right.
\]%
\begin{equation}
-2\frac{q_{\alpha\bot}}{\vec{q}^{\,\,2}}\frac{1}{\vec{k}^{\,\,2}}\left\{
-e^{-i\,\vec{k}\,\vec{z}_{10}-i\,\vec{q}\,\vec{z}_{30}}\ln\frac{\vec
{k}^{\,\,2}}{(\vec{q}+\vec{k})^{2}}-e^{-i\vec{q}\vec{z}_{30}+i\,\vec{k}%
\,\vec{z}_{31}}\ln\frac{\vec{q}^{\,\,2}}{\vec{k}^{\,\,2}}\right\}
\end{equation}%
\begin{equation}
+e^{i(\vec{q}\vec{z}_{03}+\vec{k}\vec{z}_{01})}\left.  \left\{  -\frac
{q_{\alpha\bot}}{\vec{q}^{\,\,2}\vec{k}^{\,\,2}}\ln\frac{\vec{q}^{\,\,2}%
}{(\vec{q}+\vec{k})^{2}}+\frac{k_{\alpha\bot}}{\vec{q}^{\,\,2}\vec{k}^{\,\,2}%
}\ln\frac{\vec{k}^{2}}{(\vec{q}+\vec{k})^{2}}-\frac{2(\vec{q}\vec
{k})(k+q)_{\alpha\bot}}{\vec{q}^{\,\,2}\vec{k}^{\,\,2}(\vec{q}+\vec{k})^{2}%
}\ln\frac{\vec{k}^{2}}{\vec{q}^{\,\,2}}\right\}  \right)  .
\end{equation}
Then we Furier transform this expression introducing $\vec{x}$ integration via%
\[
\langle K_{NLO}\otimes B_{123}^{\eta}\rangle|_{11+9+10+(9+10)(3\leftrightarrow
1)}=2g^{4}f^{a^{\prime}b^{\prime}c^{\prime}}(U_{1}t^{a})\cdot(t^{b}U_{2}%
)\cdot(U_{3}t^{c})\int d\vec{z}_{0}U_{0}^{bb^{\prime}}\frac{(z_{20})_{\bot
}^{\alpha}}{\left(  2\pi\right)  ^{7}\vec{z}_{20}^{\,\,2}}%
\]%
\[
\times\int d\vec{k}\int d\vec{q}\int\frac{d\vec{x}d\vec{p}}{\left(
2\pi\right)  ^{2}}\frac{p_{\bot\alpha}}{\vec{p}^{\,\,2}}e^{i\,(\vec{k}%
\,-\vec{p})\vec{z}_{1x}}\left(  \frac{-2}{\vec{q}^{\,\,2}}\left\{
e^{-i\,\vec{k}\,\vec{z}_{10}-i\,\vec{q}\,\vec{z}_{30}}\ln\frac{\vec{q}%
^{\,\,2}}{(\vec{q}+\vec{k})^{2}}+e^{-i\vec{k}\vec{z}_{10}-i\,\vec{q}\,\vec
{z}_{31}}\ln\frac{\vec{k}^{2}}{\vec{q}^{\,\,2}}\right\}  \right.
\]%
\begin{equation}
\left.  \times e^{i(\vec{q}\vec{z}_{03}+\vec{k}\vec{z}_{01})}\left\{
-\frac{1}{\vec{k}^{\,\,2}}\ln\frac{\vec{q}^{\,\,2}}{(\vec{q}+\vec{k})^{2}%
}e^{i\vec{z}_{x3}(\vec{p}-\vec{q})}-\frac{(\vec{q}\vec{k})}{\vec{q}%
^{\,\,2}\vec{k}^{\,\,2}}\ln\frac{\vec{k}^{2}}{\vec{q}^{\,\,2}}e^{i\vec{z}%
_{0x}(\vec{p}-\vec{q}-\vec{k})}\right\}  -\left(  \vec{k}\leftrightarrow
\vec{q}\right)  \right)  .
\end{equation}
All the integrals necessary to take Fourier transform one can find in
Appendices A and B of \cite{Fadin:2007de}. As a result%
\[
\langle K_{NLO}\otimes B_{123}^{\eta}\rangle|_{11+9+10+(9+10)(3\leftrightarrow
1)}=if^{abc}g^{4}(U_{1}t^{a})\cdot(t^{b^{\prime}}U_{2})\cdot(U_{3}t^{c}%
)\int\frac{U_{0}^{b^{\prime}b}d\vec{z}_{0}}{\left(  2\pi\right)  ^{6}\vec
{z}_{20}^{\,\,\,2}}%
\]%
\[
\times\left[  2\int d\vec{x}\left\{  \frac{\left(  \vec{z}_{1x}\vec{z}%
_{20}\right)  }{\vec{z}_{x1}^{\,\,2}}\frac{1}{\vec{z}_{x0}{}^{2}}\ln\frac
{\vec{z}_{x3}^{\,\,2}{}}{\vec{z}_{03}^{\,\,2}}-2\frac{\left(  \vec{z}_{0x}%
\vec{z}_{20}\right)  }{\vec{z}_{0x}^{\,\,2}}\frac{(\vec{z}_{x3}\vec{z}_{x1}%
)}{\vec{z}_{x3}^{\,\,2}\vec{z}_{x1}^{\,\,2}}\ln\vec{z}_{x3}^{\,\,2}{}\right\}
\right.
\]%
\[
\left.  +2\pi\frac{\left(  \vec{z}_{10}\vec{z}_{20}\right)  }{\vec{z}%
_{01}^{\,\,2}}\ln\frac{\vec{z}_{03}^{\,\,2}}{\vec{z}_{13}^{\,\,2}}\ln
\frac{\vec{z}_{10}^{\,\,4}}{\vec{z}_{03}^{\,\,2}\vec{z}_{13}^{\,\,2}%
}-(1\leftrightarrow3)\right]  .
\]
This integral can be calculated changing the variables to $\rho=\left\vert
\vec{z}_{x3}\right\vert ,t=e^{i\phi_{\vec{z}_{x3}}}$ and integrating with
respect to $t$ via residues. After that one has dilogarithmic integrals which
can be combined to
\[
\langle K_{NLO}\otimes B_{123}^{\eta}\rangle|_{11+9+10+(9+10)(3\leftrightarrow
1)}=if^{abc}g^{4}(U_{1}t^{a})\cdot(t^{b^{\prime}}U_{2})\cdot(U_{3}t^{c}%
)\int\frac{U_{0}^{b^{\prime}b}d\vec{z}_{0}}{\left(  2\pi\right)  ^{5}}%
\]%
\begin{equation}
\times\left[  \frac{\left(  \vec{z}_{10}\vec{z}_{20}\right)  }{\vec{z}%
_{10}^{\,2}\vec{z}_{20}^{\,\,\,2}}-\frac{\left(  \vec{z}_{30}\vec{z}%
_{20}\right)  }{\vec{z}_{30}^{\,2}\vec{z}_{20}^{\,\,\,2}}\right]  \ln
\frac{\vec{z}_{30}^{\,\,2}}{\vec{z}_{31}^{\,\,2}}\ln\frac{\vec{z}_{10}%
^{\,\,2}}{\vec{z}_{31}^{\,\,2}}.
\end{equation}
Therefore the sum of these diagrams and the diagrams which are the mirror
reflection of them with respect to the shockwave reads%
\[
\langle K_{NLO}^{conn}\otimes B_{123}^{\eta}\rangle|_{1g}=ig^{4}\left\{
f^{abc}(U_{1}t^{a})\cdot(t^{b^{\prime}}U_{2})\cdot(U_{3}t^{c})-f^{ab^{\prime
}c}(t^{a}U_{1})\cdot(U_{2}t^{b})\cdot(t^{c}U_{3})\right\}
\]%
\begin{equation}
\times\int\frac{U_{0}^{b^{\prime}b}d\vec{z}_{0}}{\left(  2\pi\right)  ^{5}%
}\left[  \frac{\left(  \vec{z}_{10}\vec{z}_{20}\right)  }{\vec{z}_{10}%
^{\,2}\vec{z}_{20}^{\,\,\,2}}-\frac{\left(  \vec{z}_{30}\vec{z}_{20}\right)
}{\vec{z}_{30}^{\,2}\vec{z}_{20}^{\,\,\,2}}\right]  \ln\frac{\vec{z}%
_{30}^{\,\,2}}{\vec{z}_{31}^{\,\,2}}\ln\frac{\vec{z}_{10}^{\,\,2}}{\vec
{z}_{31}^{\,\,2}}. \label{g1allUnconvolved}%
\end{equation}
Performing the convolution, one gets%
\[
\langle K_{NLO}^{conn}\otimes B_{123}^{\eta}\rangle|_{1g}=\frac{\alpha_{s}%
^{2}}{8\pi^{3}}\int d\vec{z}_{0}\left[  \frac{\left(  \vec{z}_{10}\vec{z}%
_{20}\right)  }{\vec{z}_{10}^{\,2}\vec{z}_{20}^{\,\,\,2}}-\frac{\left(
\vec{z}_{30}\vec{z}_{20}\right)  }{\vec{z}_{30}^{\,2}\vec{z}_{20}^{\,\,\,2}%
}\right]  \ln\frac{\vec{z}_{30}^{\,\,2}}{\vec{z}_{31}^{\,\,2}}\ln\frac{\vec
{z}_{10}^{\,\,2}}{\vec{z}_{31}^{\,\,2}}\left(  B_{100}B_{320}-B_{300}%
B_{210}\right)
\]%
\begin{equation}
+(2\leftrightarrow1)+(2\leftrightarrow3). \label{1gResultCoordinate}%
\end{equation}
From (\ref{g1allUnconvolved}) one can find the contribution of the
disconnected diagrams, which differ from the ones in fig. \ref{1gluoninSW}
in the attachment of the gluon in the right hand side of the diagrams, and the
ones which they go into after the mirror reflection with respect to the
shockwave. We get for the sum of (\ref{1gResultCoordinate}) and all such
diagrams
\[
\langle\tilde{K}_{NLO}\otimes B_{123}^{\eta}\rangle|_{1g}=\frac{g^{4}}{4}%
\int\frac{d\vec{z}_{0}}{\left(  2\pi\right)  ^{5}}\left[  \frac{\left(
\vec{z}_{10}\vec{z}_{20}\right)  }{\vec{z}_{10}^{\,2}\vec{z}_{20}^{\,\,\,2}%
}-\frac{\left(  \vec{z}_{30}\vec{z}_{20}\right)  }{\vec{z}_{30}^{\,2}\vec
{z}_{20}^{\,\,\,2}}\right]  \ln\frac{\vec{z}_{30}^{\,\,2}}{\vec{z}%
_{31}^{\,\,2}}\ln\frac{\vec{z}_{10}^{\,\,2}}{\vec{z}_{31}^{\,\,2}}\left(
B_{100}B_{320}-B_{300}B_{210}\right)
\]%
\[
+\frac{g^{4}}{4}\int\frac{d\vec{z}_{0}}{\left(  2\pi\right)  ^{5}}\left[
\frac{1}{\vec{z}_{10}^{\,2}}-\frac{\left(  \vec{z}_{30}\vec{z}_{10}\right)
}{\vec{z}_{30}^{\,2}\vec{z}_{10}^{\,\,\,2}}\right]  \ln\frac{\vec{z}%
_{30}^{\,\,2}}{\vec{z}_{31}^{\,\,2}}\ln\frac{\vec{z}_{10}^{\,\,2}}{\vec
{z}_{31}^{\,\,2}}\left(  B_{123}-\frac{1}{2}\left[  3B_{100}B_{320}%
+B_{300}B_{120}-B_{200}B_{130}\right]  \right)
\]%
\[
+\frac{g^{4}}{4}\int\frac{d\vec{z}_{0}}{\left(  2\pi\right)  ^{5}}\left[
\frac{\left(  \vec{z}_{10}\vec{z}_{30}\right)  }{\vec{z}_{10}^{\,2}\vec
{z}_{30}^{\,\,\,2}}-\frac{1}{\vec{z}_{30}^{\,2}}\right]  \ln\frac{\vec{z}%
_{30}^{\,\,2}}{\vec{z}_{31}^{\,\,2}}\ln\frac{\vec{z}_{10}^{\,\,2}}{\vec
{z}_{31}^{\,\,2}}\left(  \frac{1}{2}\left[  3B_{300}B_{120}+B_{100}%
B_{320}-B_{200}B_{130}\right]  -B_{123}\right)
\]%
\begin{equation}
+(2\leftrightarrow1)+(2\leftrightarrow3).
\end{equation}
If we put here $\vec{z}_{3}=\vec{z}_{2}$, we get%
\begin{equation}
\langle\tilde{K}_{NLO}\otimes tr(U_{1}U_{2}^{\dag})\rangle|_{1g}=-\frac
{\alpha_{s}^{2}}{4\pi^{3}}\int d\vec{z}_{0}\frac{\vec{z}_{12}^{\,\,2}}{\vec
{z}_{10}^{\,2}\vec{z}_{20}^{\,\,\,2}}\ln\frac{\vec{z}_{20}^{\,\,2}}{\vec
{z}_{21}^{\,\,2}}\ln\frac{\vec{z}_{10}^{\,\,2}}{\vec{z}_{21}^{\,\,2}}\left(
3tr(U_{1}U_{0}^{\dag})tr(U_{0}U_{2}^{\dag})-tr(U_{1}U_{2}^{\dag})\right)  ,
\end{equation}
which coincides with the corresponding contribution to the color dipole kernel
(see expression (99) in \cite{Balitsky:2008zza}).

\section{Linearized C-odd connected contribution in the momentum space}

In this section we linearize and Furier transform the connected part of the
kernel (\ref{2gResultCoordinate}) and (\ref{1gResultCoordinate}) for the C-odd
case. All the integrals necessary to take Fourier transform one can find in
Appendices A and B of \cite{Fadin:2007de}. We introduce the C-odd and C-even
Green functions%
\begin{equation}
B_{123}^{-}=B_{123}^{\eta}-B_{\bar{1}\bar{2}\bar{3}}^{\eta},\quad B_{123}%
^{+}=B_{123}^{\eta}+B_{\bar{1}\bar{2}\bar{3}}^{\eta}-12,
\end{equation}
where%
\begin{equation}
B_{\bar{1}\bar{2}\bar{3}}^{\eta}=U_{1}^{\dag}\cdot U_{2}^{\dag}\cdot
U_{3}^{\dag}.
\end{equation}
We start from (\ref{1gResultCoordinate}). For the C-odd case in the linear
regime we have%
\[
\langle K_{NLO}^{conn}\otimes B_{123}^{-}\rangle|_{1g}=\frac{3\alpha_{s}^{2}%
}{4\pi^{3}}\int d\vec{z}_{0}\left[  \frac{\left(  \vec{z}_{10}\vec{z}%
_{20}\right)  }{\vec{z}_{10}^{\,2}\vec{z}_{20}^{\,\,\,2}}-\frac{\left(
\vec{z}_{30}\vec{z}_{20}\right)  }{\vec{z}_{30}^{\,2}\vec{z}_{20}^{\,\,\,2}%
}\right]  \ln\frac{\vec{z}_{30}^{\,\,2}}{\vec{z}_{31}^{\,\,2}}\ln\frac{\vec
{z}_{10}^{\,\,2}}{\vec{z}_{31}^{\,\,2}}%
\]%
\begin{equation}
\times\left(  B_{023}^{-}+B_{100}^{-}-B_{120}^{-}-B_{003}^{-}\right)
+(2\leftrightarrow1)+(2\leftrightarrow3).
\end{equation}
One can rewrite it as
\begin{equation}
\langle K_{NLO}^{conn}\otimes B_{123}^{-}\rangle|_{1g}=\int d\vec
{z}_{1^{\prime}}d\vec{z}_{2^{\prime}}d\vec{z}_{3^{\prime}}K_{NLO}%
^{conn}\left(  \vec{z}_{1},\vec{z}_{2},\vec{z}_{3};\vec{z}_{1^{\prime}}%
,\vec{z}_{2^{\prime}},\vec{z}_{3^{\prime}}\right)  |_{1g}B_{1^{\prime
}2^{\prime}3^{\prime}}^{-},
\end{equation}
where%
\[
K_{NLO}^{conn}\left(  \vec{z}_{1},\vec{z}_{2},\vec{z}_{3};\vec{z}_{1^{\prime}%
},\vec{z}_{2^{\prime}},\vec{z}_{3^{\prime}}\right)  |_{1g}=\frac{3\alpha
_{s}^{2}}{4\pi^{3}}\int d\vec{z}_{0}\left[  \frac{\left(  \vec{z}_{10}\vec
{z}_{20}\right)  }{\vec{z}_{10}^{\,2}\vec{z}_{20}^{\,\,\,2}}-\frac{\left(
\vec{z}_{30}\vec{z}_{20}\right)  }{\vec{z}_{30}^{\,2}\vec{z}_{20}^{\,\,\,2}%
}\right]  \ln\frac{\vec{z}_{30}^{\,\,2}}{\vec{z}_{31}^{\,\,2}}\ln\frac{\vec
{z}_{10}^{\,\,2}}{\vec{z}_{31}^{\,\,2}}%
\]%
\[
\times\left(  \delta\left(  \vec{z}_{22^{\prime}}\right)  -\delta\left(
\vec{z}_{02^{\prime}}\right)  \right)  \left(  \delta\left(  \vec
{z}_{01^{\prime}}\right)  \delta\left(  \vec{z}_{33^{\prime}}\right)
-\delta\left(  \vec{z}_{11^{\prime}}\right)  \delta\left(  \vec{z}%
_{03^{\prime}}\right)  \right)
\]%
\begin{equation}
+(2\leftrightarrow1)+(2\leftrightarrow3).
\end{equation}
The kernel in the momentum representation reads%
\[
K\left(  \vec{q}_{1},\vec{q}_{2},\vec{q}_{3};\vec{q}_{1^{\prime}},\vec
{q}_{2^{\prime}},\vec{q}_{3^{\prime}}\right)  =\int\frac{d\vec{z}_{1}}{2\pi
}\frac{d\vec{z}_{2}}{2\pi}\frac{d\vec{z}_{3}}{2\pi}\frac{d\vec{z}_{1}^{\prime
}}{2\pi}\frac{d\vec{z}_{2}^{\prime}}{2\pi}\frac{d\vec{z}_{3}^{\prime}}{2\pi
}e^{-i[\vec{q}_{1}\vec{z}_{1}+\vec{q}_{2}\vec{z}_{2}+\vec{q}_{3}\vec{z}%
_{3}-\vec{q}_{1}^{\prime}\vec{z}_{1}^{\prime}-\vec{q}_{2}^{\prime}\vec{z}%
_{2}^{\prime}-\vec{q}_{3}^{\prime}\vec{z}_{3}^{\prime}]}%
\]%
\begin{equation}
\times K\left(  \vec{z}_{1},\vec{z}_{2},\vec{z}_{3};\vec{z}_{1^{\prime}}%
,\vec{z}_{2^{\prime}},\vec{z}_{3^{\prime}}\right)  . \label{Kmom}%
\end{equation}
We have%
\[
K_{NLO}^{conn}\left(  \vec{q}_{1},\vec{q}_{2},\vec{q}_{3};\vec{q}_{1^{\prime}%
},\vec{q}_{2^{\prime}},\vec{q}_{3^{\prime}}\right)  |_{1g}=\frac{3\alpha
_{s}^{2}}{4\pi^{3}\left(  2\pi\right)  }\delta\left(  \vec{q}_{11^{\prime}%
}+\vec{q}_{22^{\prime}}+\vec{q}_{33^{\prime}}\right)  \left[  \frac{\vec
{q}_{2}}{\vec{q}_{2}^{\,\,2}}-\frac{\vec{q}_{22^{\prime}}}{\vec{q}%
_{22^{\prime}}^{\,\,2}}\right]
\]%
\[
\times\left\{  -\left[  \frac{1}{\vec{q}_{33^{\prime}}^{\,\,2}}\frac{\partial
}{\partial\vec{q}_{1}}-\frac{1}{\vec{q}_{1}^{\,\,2}}\frac{\partial}%
{\partial\vec{q}_{3}}\right]  \ln\frac{\vec{q}_{33^{\prime}}^{\,\,2}}{\left(
\vec{q}_{1}+\vec{q}_{33^{\prime}}\right)  ^{2}}\ln\frac{\vec{q}_{1}^{\,\,2}%
}{\left(  \vec{q}_{1}+\vec{q}_{33^{\prime}}\right)  ^{2}}\right.
\]%
\[
+\left.  \left[  \frac{1}{\vec{q}_{3}^{\,\,2}}\frac{\partial}{\partial\vec
{q}_{1}}-\frac{1}{\vec{q}_{11^{\prime}}^{\,\,2}}\frac{\partial}{\partial
\vec{q}_{3}}\right]  \ln\frac{\vec{q}_{3}^{\,\,2}}{\left(  \vec{q}%
_{11^{\prime}}+\vec{q}_{3}\right)  ^{2}}\ln\frac{\vec{q}_{11^{\prime}}%
^{\,\,2}}{\left(  \vec{q}_{11^{\prime}}+\vec{q}_{3}\right)  ^{2}}\right\}
\]%
\begin{equation}
+(2\leftrightarrow1)+(2\leftrightarrow3). \label{1gResultMom}%
\end{equation}
To deal with (\ref{2gResultCoordinate}) one has to linearize the color
structure first. It reads%
\[
M=(U_{2}U_{0}^{\dag}U_{1})\cdot U_{4}\cdot(U_{0}U_{4}^{\dag}U_{3})+(U_{1}%
U_{0}^{\dag}U_{2})\cdot U_{4}\cdot(U_{3}U_{4}^{\dag}U_{0})
\]%
\[
=U_{1}\cdot U_{4}\cdot(U_{3}U_{4}^{\dag}U_{0}+U_{0}U_{4}^{\dag}U_{3}%
)+(U_{1}U_{0}^{\dag}U_{2}+U_{2}U_{0}^{\dag}U_{1})\cdot U_{4}\cdot U_{0}%
\]%
\[
+(U_{2}-U_{0})\cdot U_{4}\cdot((U_{0}-U_{4})U_{4}^{\dag}(U_{3}-U_{4}%
)+(U_{3}-U_{4})U_{4}^{\dag}(U_{0}-U_{4}))
\]%
\[
+((U_{1}-U_{0})U_{0}^{\dag}(U_{2}-U_{0}))\cdot U_{4}\cdot((U_{3}-U_{4}%
)U_{4}^{\dag}U_{0})
\]%
\[
+((U_{2}-U_{0})U_{0}^{\dag}(U_{1}-U_{0}))\cdot U_{4}\cdot(U_{0}U_{4}^{\dag
}(U_{3}-U_{4}))
\]%
\begin{equation}
+2(U_{2}-U_{0})\cdot U_{4}\cdot(U_{3}-U_{4})-2U_{1}\cdot U_{4}\cdot U_{0}.
\end{equation}
Then we can use SU(3) identity (\ref{IDENTITY})\ and take into account that in
the 3-gluon approximation%
\[
(U_{2}-U_{0})\cdot U_{4}\cdot((U_{0}-U_{4})U_{4}^{\dag}(U_{3}-U_{4}%
)+(U_{3}-U_{4})U_{4}^{\dag}(U_{0}-U_{4}))
\]%
\[
=(U_{2}-U_{0})\cdot E\cdot((U_{0}-U_{4})(U_{3}-U_{4})+(U_{3}-U_{4}%
)(U_{0}-U_{4}))
\]%
\begin{equation}
=-(U_{2}-U_{0})\cdot(U_{0}-U_{4})\cdot(U_{3}-U_{4}),
\end{equation}
and%
\[
((U_{1}-U_{0})U_{0}^{\dag}(U_{2}-U_{0}))\cdot U_{4}\cdot((U_{3}-U_{4}%
)U_{4}^{\dag}U_{0})
\]%
\[
+((U_{2}-U_{0})U_{0}^{\dag}(U_{1}-U_{0}))\cdot U_{4}\cdot(U_{0}U_{4}^{\dag
}(U_{3}-U_{4}))
\]%
\[
=((U_{1}-U_{0})(U_{2}-U_{0})+(U_{2}-U_{0})(U_{1}-U_{0}))\cdot E\cdot
(U_{3}-U_{4})
\]%
\begin{equation}
=-(U_{2}-U_{0})\cdot(U_{1}-U_{0})\cdot(U_{3}-U_{4}).
\end{equation}
Then in the 3-gluon approximation%
\[
M-M^{\dag}-(1,0\leftrightarrow3,4)=3(6B_{044}^{-}+3B_{014}^{-}-3B_{043}^{-}%
\]%
\begin{equation}
+2B_{200}^{-}-2B_{244}^{-}+2B_{243}^{-}-2B_{210}^{-}+B_{314}^{-}-B_{310}%
^{-}-B_{144}^{-}+B_{300}^{-}),
\end{equation}
and (\ref{2gResultCoordinate}) reads%
\[
\langle K_{NLO}^{conn}\otimes B_{123}^{\eta}\rangle|_{2g}=\frac{3\alpha
_{s}^{2}}{4\pi^{4}}\int d\vec{z}_{0}\int d\vec{z}_{4}\left\{  3B_{044}%
^{-}-3B_{004}^{-}+3B_{104}^{-}-3B_{043}^{-}\right.
\]%
\[
\left.  +2B_{423}^{-}-2B_{120}^{-}+2B_{020}^{-}-2B_{424}^{-}+B_{143}%
^{-}-B_{103}^{-}+B_{003}^{-}-B_{144}^{-}\right\}
\]%
\[
\times\left[  \frac{1}{2\vec{z}_{04}^{\,\,2}}\frac{\left(  \vec{z}_{10}\vec
{z}_{34}\right)  }{\vec{z}_{10}^{\,\,2}\vec{z}_{34}^{\,\,2}}+\frac{(\vec
{z}_{10}\vec{z}_{40})}{\vec{z}_{10}^{\,\,2}\vec{z}_{40}^{\,\,2}}\frac{(\vec
{z}_{24}\vec{z}_{34})}{\vec{z}_{24}^{\,\,2}\vec{z}_{34}^{\,\,2}}+\frac
{(\vec{z}_{04}\vec{z}_{34})}{\vec{z}_{04}^{\,\,2}\vec{z}_{34}^{\,\,2}}%
\frac{(\vec{z}_{10}\vec{z}_{20})}{\vec{z}_{10}^{\,\,2}\vec{z}_{20}^{\,\,2}%
}-\frac{\left(  \vec{z}_{20}\vec{z}_{10}\right)  }{\vec{z}_{02}{}^{2}\vec
{z}_{01}{}^{2}}\frac{\left(  \vec{z}_{24}\vec{z}_{34}\right)  }{\vec{z}%
_{24}^{\,\,2}\vec{z}_{34}{}^{2}}\right]  \ln\frac{\vec{z}_{02}^{\,\,2}}%
{\vec{z}_{24}^{\,\,2}}%
\]%
\begin{equation}
+(2\leftrightarrow1)+(2\leftrightarrow3).
\end{equation}
Therefore
\[
K_{NLO}^{conn}\left(  \vec{z}_{1},\vec{z}_{2},\vec{z}_{3};\vec{z}_{1^{\prime}%
},\vec{z}_{2^{\prime}},\vec{z}_{3^{\prime}}\right)  |_{2g}=\frac{3\alpha
_{s}^{2}}{4\pi^{4}}\int d\vec{z}_{0}\int d\vec{z}_{4}\ln\frac{\vec{z}%
_{02}^{\,\,2}}{\vec{z}_{24}^{\,\,2}}%
\]%
\[
\times\left[  \frac{1}{2\vec{z}_{04}^{\,\,2}}\frac{\left(  \vec{z}_{10}\vec
{z}_{34}\right)  }{\vec{z}_{10}^{\,\,2}\vec{z}_{34}^{\,\,2}}+\frac{(\vec
{z}_{10}\vec{z}_{40})}{\vec{z}_{10}^{\,\,2}\vec{z}_{40}^{\,\,2}}\frac{(\vec
{z}_{24}\vec{z}_{34})}{\vec{z}_{24}^{\,\,2}\vec{z}_{34}^{\,\,2}}+\frac
{(\vec{z}_{04}\vec{z}_{34})}{\vec{z}_{04}^{\,\,2}\vec{z}_{34}^{\,\,2}}%
\frac{(\vec{z}_{10}\vec{z}_{20})}{\vec{z}_{10}^{\,\,2}\vec{z}_{20}^{\,\,2}%
}-\frac{\left(  \vec{z}_{20}\vec{z}_{10}\right)  }{\vec{z}_{02}{}^{2}\vec
{z}_{01}{}^{2}}\frac{\left(  \vec{z}_{24}\vec{z}_{34}\right)  }{\vec{z}%
_{24}^{\,\,2}\vec{z}_{34}{}^{2}}\right]
\]%
\[
\times\left\{  \frac{{}}{{}}\left(  \delta\left(  \vec{z}_{01^{\prime}%
}\right)  -\delta\left(  \vec{z}_{11^{\prime}}\right)  \right)  \left(
\delta\left(  \vec{z}_{33^{\prime}}\right)  \delta\left(  \vec{z}_{02^{\prime
}}\right)  +2\delta\left(  \vec{z}_{22^{\prime}}\right)  \delta\left(  \vec
{z}_{03^{\prime}}\right)  -3\delta\left(  \vec{z}_{43^{\prime}}\right)
\delta\left(  \vec{z}_{02^{\prime}}\right)  \right)  \right.
\]%
\[
-\left.  \frac{{}}{{}}\left(  \delta\left(  \vec{z}_{41^{\prime}}\right)
-\delta\left(  \vec{z}_{31^{\prime}}\right)  \right)  \left(  \delta\left(
\vec{z}_{13^{\prime}}\right)  \delta\left(  \vec{z}_{42^{\prime}}\right)
+2\delta\left(  \vec{z}_{22^{\prime}}\right)  \delta\left(  \vec
{z}_{43^{\prime}}\right)  -3\delta\left(  \vec{z}_{03^{\prime}}\right)
\delta\left(  \vec{z}_{42^{\prime}}\right)  \right)  \right\}
\]%
\begin{equation}
+(2\leftrightarrow1)+(2\leftrightarrow3).
\end{equation}
Hence this contribution to the kernel in the momentum representation
(\ref{Kmom}) reads%
\[
K_{NLO}^{conn}\left(  \vec{q}_{1},\vec{q}_{2},\vec{q}_{3};\vec{q}_{1^{\prime}%
},\vec{q}_{2^{\prime}},\vec{q}_{3^{\prime}}\right)  |_{2g}=\left\{  \left(
2\pi\right)  ^{2}\delta\left(  \vec{q}_{11^{\prime}}+\vec{q}_{22^{\prime}%
}+\vec{q}_{33^{\prime}}\right)  \frac{3\alpha_{s}^{2}}{4\pi^{4}}\int
\frac{d\vec{z}_{10}}{\left(  2\pi\right)  ^{2}}\frac{d\vec{z}_{20}}{\left(
2\pi\right)  ^{2}}\frac{d\vec{z}_{34}}{\left(  2\pi\right)  ^{2}}\right.
\]%
\[
\times\int d\vec{z}_{40}\ln\frac{\vec{z}_{20}^{\,\,2}}{\left(  \vec{z}%
_{20}-\vec{z}_{40}\right)  ^{2}}\frac{z_{10}^{i}}{\vec{z}_{10}^{\,\,2}}%
\frac{z_{34}^{j}}{\vec{z}_{34}^{\,\,2}}\left[  \frac{\delta^{ij}}{2\vec
{z}_{40}^{\,\,2}}+\left\{  \frac{z_{40}^{i}}{\vec{z}_{40}^{\,\,2}}%
-\frac{z_{20}^{i}}{\vec{z}_{20}{}^{2}}\right\}  \frac{\left(  z_{20}%
-z_{40}\right)  ^{j}}{\left(  \vec{z}_{20}-\vec{z}_{40}\right)  ^{2}}%
-\frac{z_{40}^{j}}{\vec{z}_{40}^{\,\,2}}\frac{z_{20}^{i}}{\vec{z}_{20}%
^{\,\,2}}\right]
\]%
\[
\times\left(  (e^{-i[\vec{q}_{2}\vec{z}_{20}+\vec{q}_{33^{\prime}}\vec{z}%
_{34}+\vec{q}_{33^{\prime}}\vec{z}_{40}]}+2e^{-i[\vec{q}_{22^{\prime}}\vec
{z}_{20}+\vec{q}_{3}\vec{z}_{34}+\vec{q}_{3}\vec{z}_{40}]}-3e^{-i[\vec{q}%
_{2}\vec{z}_{20}+\vec{q}_{3}\vec{z}_{34}+\vec{q}_{33^{\prime}}\vec{z}_{40}%
]})\right)
\]%
\begin{equation}
\left.  \times(e^{-i\vec{q}_{1}\vec{z}_{10}}-e^{-i\vec{q}_{11^{\prime}}\vec
{z}_{10}})+\left(  1\leftrightarrow3,0\leftrightarrow4,1^{\prime
}\leftrightarrow3^{\prime}\right)  \frac{{}}{{}}\right\}  +(2\leftrightarrow
1)+(2\leftrightarrow3).
\end{equation}
We can rewrite it as\qquad%
\[
K_{NLO}^{conn}\left(  \vec{q}_{1},\vec{q}_{2},\vec{q}_{3};\vec{q}_{1^{\prime}%
},\vec{q}_{2^{\prime}},\vec{q}_{3^{\prime}}\right)  |_{2g}=\left\{
\delta\left(  \vec{q}_{11^{\prime}}+\vec{q}_{22^{\prime}}+\vec{q}_{33^{\prime
}}\right)  \int\frac{d\vec{z}_{20}}{\left(  2\pi\right)  ^{2}}\frac
{3\alpha_{s}^{2}}{4\pi^{4}}\int d\vec{z}_{40}\right.
\]%
\[
\times\ln\frac{\vec{z}_{20}^{\,\,2}}{\left(  \vec{z}_{20}-\vec{z}_{40}\right)
^{2}}\left(  \frac{q_{11^{\prime}}^{i}}{\vec{q}_{11^{\prime}}^{\,\,2}}%
-\frac{q_{1}^{i}}{\vec{q}_{1}^{\,\,2}}\right)  \left[  \frac{\delta^{ij}%
}{2\vec{z}_{40}^{\,\,2}}+\left\{  \frac{z_{40}^{i}}{\vec{z}_{40}^{\,\,2}%
}-\frac{z_{20}^{i}}{\vec{z}_{20}{}^{2}}\right\}  \frac{\left(  z_{20}%
-z_{40}\right)  ^{j}}{\left(  \vec{z}_{20}-\vec{z}_{40}\right)  ^{2}}%
-\frac{z_{40}^{j}}{\vec{z}_{40}^{\,\,2}}\frac{z_{20}^{i}}{\vec{z}_{20}%
^{\,\,2}}\right]
\]%
\[
\times\left(  \left(  \frac{q_{33^{\prime}}^{j}}{\vec{q}_{33^{\prime}}%
^{\,\,2}}-\frac{q_{3}^{j}}{\vec{q}_{3}^{\,\,2}}\right)  e^{-i[\vec{q}_{2}%
\vec{z}_{20}+\vec{q}_{33^{\prime}}\vec{z}_{40}]}+2\frac{q_{3}^{j}}{\vec{q}%
_{3}^{\,\,2}}\left(  e^{-i[\vec{q}_{22^{\prime}}\vec{z}_{20}+\vec{q}_{3}%
\vec{z}_{40}]}-e^{-i[\vec{q}_{2}\vec{z}_{20}+\vec{q}_{33^{\prime}}\vec{z}%
_{40}]}\right)  \right)
\]%
\begin{equation}
\left.  +\left(  1\leftrightarrow3,0\leftrightarrow4,1^{\prime}\leftrightarrow
3^{\prime}\right)  \frac{{}}{{}}\right\}  +(2\leftrightarrow
1)+(2\leftrightarrow3).
\end{equation}
Finally%
\[
K_{NLO}^{conn}\left(  \vec{q}_{1},\vec{q}_{2},\vec{q}_{3};\vec{q}_{1^{\prime}%
},\vec{q}_{2^{\prime}},\vec{q}_{3^{\prime}}\right)  |_{2g}=\delta\left(
\vec{q}_{11^{\prime}}+\vec{q}_{22^{\prime}}+\vec{q}_{33^{\prime}}\right)
\frac{3\alpha_{s}^{2}}{4\pi^{4}}%
\]%
\[
\times\left\{  \left(  \frac{q_{11^{\prime}}^{i}}{\vec{q}_{11^{\prime}%
}^{\,\,2}}-\frac{q_{1}^{i}}{\vec{q}_{1}^{\,\,2}}\right)  \left(  \left(
\frac{q_{33^{\prime}}^{j}}{\vec{q}_{33^{\prime}}^{\,\,2}}-\frac{q_{3}^{j}%
}{\vec{q}_{3}^{\,\,2}}\right)  A^{ij}\left(  q_{2},q_{33^{\prime}}\right)
+2\frac{q_{3}^{j}}{\vec{q}_{3}^{\,\,2}}\left(  A^{ij}\left(  q_{22^{\prime}%
},q_{3}\right)  -A^{ij}\left(  q_{2},q_{33^{\prime}}\right)  \right)  \right)
\right.
\]%
\[
+\left.  \left(  \frac{q_{33^{\prime}}^{i}}{\vec{q}_{33^{\prime}}^{\,\,2}%
}-\frac{q_{3}^{i}}{\vec{q}_{3}^{\,\,2}}\right)  \left(  \left(  \frac
{q_{11^{\prime}}^{j}}{\vec{q}_{11^{\prime}}^{\,\,2}}-\frac{q_{1}^{j}}{\vec
{q}_{1}^{\,\,2}}\right)  A^{ij}\left(  q_{2},q_{11^{\prime}}\right)
+2\frac{q_{1}^{j}}{\vec{q}_{1}^{\,\,2}}\left(  A^{ij}\left(  q_{22^{\prime}%
},q_{1}\right)  -A^{ij}\left(  q_{2},q_{11^{\prime}}\right)  \right)  \right)
\right\}
\]%
\begin{equation}
+(2,2^{\prime}\leftrightarrow1,1^{\prime})+(2,2^{\prime}\leftrightarrow
3,3^{\prime}), \label{2gResultMom}%
\end{equation}
where%
\begin{equation}
A^{ij}\left(  q_{2},q_{33^{\prime}}\right)  =\ln\frac{\vec{q}_{33^{\prime}%
}^{\,\,2}}{\left(  \vec{q}_{33^{\prime}}+\vec{q}_{2}\right)  ^{2}}\left\{
\frac{\delta^{ij}}{2\vec{q}_{2}^{\,\,2}}+\frac{q_{2}^{i}q_{33^{\prime}}^{j}%
}{\vec{q}_{2}^{\,\,2}\vec{q}_{33^{\prime}}^{\,\,2}}-\frac{\left(
q_{33^{\prime}}+q_{2}\right)  ^{i}q_{2}^{j}}{\left(  \vec{q}_{33^{\prime}%
}+\vec{q}_{2}\right)  ^{2}\vec{q}_{2}^{\,\,2}}-\frac{q_{33^{\prime}}%
^{j}(q_{33^{\prime}}+q_{2})^{i}}{\vec{q}_{33^{\prime}}^{\,\,2}\left(  \vec
{q}_{33^{\prime}}+\vec{q}_{2}\right)  ^{2}}\right\}  .
\end{equation}

\section{Conclusion}

The connected part of the NLO kernel for 3QWL operator has been calculated
here within Balitsky high energy operator expansion formalism
\cite{Balitsky:2008zza}. The result consists of two parts
(\ref{1gResultCoordinate}) and (\ref{2gResultCoordinate}), which represent the
contribution of diagrams with 1 and 2 gluon states crossing the shockwave. The
momentum representation of these contributions in the linear limit for C-odd
case is given in (\ref{1gResultMom}) and (\ref{2gResultMom}). Comparing these
expressions with the connected $3\rightarrow3$ contribution to odderon kernel,
obtained in \cite{Bartels:2012sw} in the momentum representation, one can see
that they do not coincide. This fact indicates that there should be an
equivalence transformation connecting the whole kernels obtained in the high
energy operator expansion formalism and in the formalism based on
reggeization. Moreover, the construction of a matrix element of a gauge
invariant operator in the momentum representation from its Mobius form in the
coordinate space consists of two steps \cite{Fadin:2011jg}. First one does the
Fourier transform and then adds to the result such terms that restore its
gauge invariance but vanish after the convolution with the colorless impact
factors. This procedure is to be applied to the whole kernel after its calculation.

\acknowledgments
I am grateful to prof. V. S. Fadin for proposing this work and to Abdus Salam
ICTP and Nordita for invitation, hospitality and support. I also thank the
Dynasty foundation for a travel grant to ``Beyond the LHC'' program in
Nordita. The study was supported by the Ministry of education of Russia
projects 14.B37.21.1181 and 8408 and by the Russian Fund for Basic Research
grants 12-02-31086, 13-02-01023 and 12-02-33140 and by president grant MK-525.2013.2.

\end{document}